\documentclass[times]{preprint}

\usepackage{shortCut}
\title{Multi-stage high order semi-Lagrangian schemes \\
for incompressible flows in Cartesian geometries}
\author{Alexandre Cameron, Rapha\"{e}l Raynaud\footnotemark[2], Emmanuel Dormy\footnotemark[4]}

\begin{document}

\begin{abstract}
Efficient transport algorithms are essential to the numerical resolution 
of incompressible fluid flow problems.
Semi-Lagrangian methods are widely used in grid based methods to achieve this aim.
The accuracy of the interpolation strategy then determines the properties of the scheme.
We introduce a simple multi-stage procedure which can easily be used to increase the order 
of accuracy of a code based on multi-linear interpolations. This approach is an extension
of a corrective algorithm introduced by Dupont \& Liu (2003, 2007).
This multi-stage procedure can be easily implemented in existing parallel codes using a domain 
decomposition strategy, as the communications pattern is identical to that of the multi-linear scheme.
We show how a combination of a forward and backward error correction can provide a 
third-order accurate scheme, thus significantly reducing diffusive effects 
while retaining a non-dispersive leading error term.

\end{abstract}

%\keywords{Transport in fluids, Finite volumes, Finite differences, Semi-Lagrangian}

\maketitle

\footnotetext[2]{Current address: School of Astronomy, 
Institute for Research in Fundamental Sciences
(IPM), 19395-5531, Tehran, Iran.}
\footnotetext[4]{Current address: D\'epartement de Math\'ematiques et Applications, 
CNRS UMR-8553,  \'{E}cole Normale Sup\'{e}rieure, 45 rue d'Ulm, 75005 Paris, France.}

%%%%% Intro %%%%%
\section{Introduction}
\label{sec:intro}
%%%% INTRO SEMI-LAG
Semi-Lagrangian methods offer an efficient and widely used approach to model
advection dominated problems. 
Initially introduced in atmospheric and weather models
\cite{robert_stable_1981, robert_semi-lagrangian_1982}, these methods are now
widely used in all fields of fluid mechanics 
\cite{staniforth_semi-lagrangian_1991, oliveira_comparison_1995,durran_numerical_1999}.
They have found a wide range of application in computational fluid
dynamics. 
These methods have triggered a wide variety of schemes, including
spline interpolation methods \cite{knorr_representation_1980, 
shoucri_numerical_1983,zerroukat_application_2007}, finite element 
WENO algorithms \cite{liu_weighted_1994,qiu_conservative_2011,
huang_eulerianlagrangian_2012} or CIP methods \cite{
nakamura_exactly_2001,xiao_completely_2001}. 
Considerable development has also been achieved in 
application to hyperbolic problems (e.g. compressible 
hydrodynamics \cite{lentine_unconditionally_2011}, Vlasov equation 
\cite{sonnendrucker_semi-lagrangian_1999}) and fall out of the scope of this paper.

%%%% Semi-Lag
Semi-Lagrangian methods involve an advected field $\Phi$, 
following the characteristics backward in time. 
The procedure requires the estimation of field values that do not lie on the computational grid. 
Semi-Lagrangian methods therefore rely on an interpolation of 
$\Phi (t-\Delta t, \xvec - \U \Delta t ) $, which in general is not a known quantity on the discrete grid. 

%%%% PARALLEL
Because of their local nature, low order semi-Lagrangian methods perform remarkably well on
massively parallel computers \cite{liu_real-time_2004, wu_improved_2004}.
Limitations occur with high-order interpolation methods. As the width 
of the stencil increases, the locality of the scheme is reduced and the resulting 
schemes require larger communications stencils.
When the interpolation strategy is simple, multi-linear in the case of the
\CIR{} scheme \cite{courant_solution_1952},  
the scheme is local and the computational cost is small.
If the interpolation stencil is not localized near the computational point, but near
the point where the interpolated value must be reconstructed, one can
show that the method is then unconditionally stable, in
the case of a uniform and steady velocity field 
\cite{leonard_stability_2002}. Such
schemes are however prone to large inter-process communations, and are not
unconditionally stable for general flows.

%%% MULTI-STAGE
Dupont \textit{et al.} \cite{dupont2003, dupont2007, kim2007} 
introduced two new corrective algorithms: ``Forward Error Correction'' (here denoted \FEC ) 
and ``Backward Error Correction'' (here denoted \BEC). These algorithms
take advantage of the reversibility of the advection equation to
improve the order of most semi-Lagrangian schemes by using multiple 
calls of an initial advection scheme. The resulting schemes yield an enhanced
accuracy. In that sense, they are built with a similar spirit
to the predictor-corrector method \cite{butcher_numerical_2008} 
or the MacCormack scheme \cite{maccormack_effect_2003}.

Here we introduce a new scheme following this methodology, and 
thus extend this approach to third order accuracy.

\section{Multi-stage approaches}
A possible strategy to increase the order of Semi-Lagrangian schemes is to use
higher order interpolation formula e.g. \cite{celledoni_high_2015}. This has the drawback of
relying on a wider stencil, which requires larger communication patterns
on a distributed memory computer. Another significant issue with wider
stencils is the handling of boundary conditions. 

Equation \eqref{eq:intro:adv} models the advection of a passive scalar $\Phi$ 
by a velocity field $\U$,
\begin{align}
	\Dt \Phi \equiv
	\left[ \partial_t + (\U \cdot \grad ) \right] \Phi =0 \, .
	\label{eq:intro:adv}
\end{align}
The Lagrangian derivative in \eqref{eq:intro:adv} is usually defined
as the limit, following the characteristic, of
\begin{align}
	D_t \Phi = \lim _{\Delta t \to 0} \dfrac{ \Phi (t, \xvec) - 
	\Phi (t-\Delta t, \xvec - \U \Delta t ) }{\Delta t} \,.
\end{align}	
Semi-Lagrangian methods rely on this expression to discretize the advective operator $D_t \Phi $ 
instead of expanding the sum in a temporal term $\partial_t \Phi $ 
and an advective term $( \U \cdot \grad) \Phi $, as in \eqref{eq:intro:adv}.
The semi-Lagrangian discretisation of \eqref{eq:intro:adv} therefore introduces 
an interpolation operator
\(
	L _{\U} \left[ \Phi ^n \right] = \widetilde \Phi ^n (\xvec - \U \Delta t) \, ,
\)
where $\widetilde \Phi$ denotes the interpolated value away from the grid 
points.

A strategy introduced by Dupont \textit{et al.} \cite{dupont2003}
to increase the order of a semi-Lagrangian scheme, without requiring the use of high-order 
interpolation formula, is based on two consecutive calls to the interpolation operator,   
the second call involving the reversed flow.
This method is known as the ``Forward Error Correction'' \cite{dupont2003}. 
The advantages of this procedure over the above high order schemes rely both on the 
accurate implementation of boundary conditions and on the limited communication stencil.
The Forward Error Correction scheme is constructed as
\begin{align}
	\bar{\Phi} &\equiv L_{-\U}\left[ L_{\U}\left[ \Phi ^n\right] \right] \, , \\
	FEC \left[\Phi ^n\right] & \equiv 
	L_{\U}\left[\Phi ^{n} \right]+ \left(\Phi ^n - \bar{\Phi}\right)/2 
	\, .
	\label{eq:intro:FEC}
\end{align}

The \FEC{} corrective algorithm has further been improved in \cite{dupont2007, kim2007}
using three calls to the interpolation operator for each time-step.
The resulting algorithm is known as the ``Backwards Error Correction'' 
(\BEC) algorithm. It is constructed using 
\begin{align}
	BEC \left[ \Phi ^n\right] \equiv 
	L_{\U} \left[ \Phi^{n} + (\Phi ^n - \bar{\Phi} )/2 \right ] 
	\,.
	\label{eq:intro:BEC}
\end{align}
Both the \FEC{} and the \BEC{} algorithms suppress the leading
order error term when the interpolation operator is irreversible.
Both the \FEC{} and the \BEC{} schemes are
free of numerical diffusion. However, they introduce numerical
dispersive effects related to their truncation errors. 

This truncation error can be advantageously used to construct a scheme free of numerical 
dispersion and characterized by a fourth order derivative 
truncation error. This is achieved for the same computational 
cost as the \BEC{} scheme. 
A new ``Combined Error Correction'' (\CEC) algorithm is introduced,
using a linear combination of the \FEC{} and \BEC{} algorithms,
	\begin{align}
		CEC \left [ \Phi \right ] 
		\equiv \cA \, FEC \left [ \Phi \right ] 
		+ \cB \, BEC \left [ \Phi \right ] \, .
      \label{eq:intro:CEC}
	\end{align}	

When the \CIR{} scheme is used as the interpolation operator,
the scheme generated by the \FEC{} algorithm is similar, in 
the Eulerian framework, to the one introduced in \cite{fromm_method_1968}.
In this case, the values of the coefficients \cA{} and \cB{} 
in \eqref{eq:intro:CEC} can be explicitly determined and the stability of the resulting schemes assessed. 
In one dimension, their expression is
	\begin{align}
		3 \, \cA = 2 - {\Delta x}/{( \vert u \vert \Delta t)} 
		\quad \text{and} \quad
		\cB =1 - \cA 
		\, ,
      \label{eq:intro:CECcoef}
	\end{align}
where $\Delta t$ denotes the time-step and $\Delta x$ the grid-step.

In one dimension of space, the \CIR{} scheme is strictly equivalent to the
Eulerian upwind scheme. It is well known \cite{courant_uber_1928, 
hirsch_numerical_2007, leveque_numerical_1992} that this scheme 
is stable for Courant-Friedrichs-Lewy (CFL) numbers smaller 
than unity and introduces diffusive errors. 
The spurious diffusive effects are directly related to the truncation error of the scheme.

The generalization to $d$-dimension must be carried
out with care. As described later, the fields can be advected one 
dimension at a time using a splitting technique similar to \cite{fromm_method_1968}.
In two or three dimensions, the interpolation can be done by applying the \CEC{} 
scheme on each direction separately.

%%%%% Body %%%%%
\section{One-dimensional algorithms}
\label{sec:algo1D}
\label{sec:1Dalgo}

	In the semi-Lagrangian formalism, the advection equation can be discretized 
	using the \CIR{} scheme \cite{courant_solution_1952}. In one dimension,
	the \CIR{} scheme has the same stencil as the Upwind scheme 
	\cite{durran_numerical_1999,
	butcher_numerical_2008, hirsch_numerical_2007}
	\begin{align}
		\Phi^{CIR}_{i}=& 
		( 1 - U_i) \Phi^n[i] + U_i 
		\Phi^n[i- s_i] \,, 
	\end{align}
	where $\Phi^{n}[i]=\Phi^{n}_{i}$ denotes the value of the 
	passive scalar at time $n \,\Delta t$ and position $i \,\Delta x$,
	$s_i = \sgn ( u_i )$ the sign of the velocity and 
	$U_i = \vert u_i \vert 	{\Delta t} / {\Delta x} $ 
	the reduced velocity with $u_i$ the velocity.
	A Von Neumann stability analysis shows that the scheme is 	
	strictly stable for $U\leq1$.
	For a constant velocity, the modified equation takes the form
	\begin{align}
		\sme{ \Big[  \partial_t \Phi + u \partial_x \Phi \Big] }{\CIR} 
		= \sme{D}{\CIR} \, \partial^2_x \Phi + ...
	   \quad \text{with} \quad
	   \sme{D}{\CIR} = \left( 1- U \right)\frac{ \vert u \vert \Delta x }{2}
	   \, .
	   \label{eq:CIRdiff}
	\end{align}
	
	The \FEC{} scheme \eqref{eq:intro:FEC} is a multi-stage 
	version of the \CIR{} scheme.
	The developed expression for the \FEC{} scheme
	requires the first nearest neighbors
	for the velocity and the second nearest neighbors for the passive scalar (see Appendix~A).	
	For a constant velocity, the expression of \FEC{} is
	\begin{align}
		\FEC[\Phi]_{i}
		= - \tfrac{1}{2} U (1- U)\Phi^n[i+ 1] 
		+ (1- U^2 )\Phi^n[i] 
		+ \tfrac{1}{2} U (1 + U) \Phi^n[i- 1]
		\, .
		\label{eq:cstUHALO}
	\end{align}
	The stability analysis of \eqref{eq:cstUHALO} provides
	the following expression for the amplification factor
	\begin{align}
		\xiF = 1 - U^2 + U^2\cos (k \Delta x ) - i U \sin( k \Delta x ) \,.
	\end{align}
	The \FEC{} scheme is stable for $U\leq 1$. The modified equation associated to this scheme is
	\begin{align}
		\sme{\Big[ \partial_t\Phi + u \partial_x\Phi \Big]}{\FEC} = 
	 	- ( 1- U^2 ) \frac{u \Delta x^2}{3!} \partial^{3}_x \Phi
	 	- 3(1- U^2) 
	 	\frac{u^2 \Delta x^2 \Delta t }{4!} \partial^{4}_x \Phi 
		+ ...
		\label{eq:final_FEC}
	\end{align}

	The \BEC{} scheme, presented in \eqref{eq:intro:BEC}, 
	is a modified version of the \CIR{} scheme using
	$\bar{\Phi}^{n}$ to correct the field before the advection step. 
	The developed expression of the \BEC{} scheme
	requires the second nearest neighbors for the velocity and third nearest neighbors for the passive scalar  (see Appeendix~A).
	To avoid using this long development, the simplified case of a constant velocity
	will be studied. 
	\begin{align}
		\label{eq:cstUBEC}
		\BEC [\Phi]_i = &
		- \frac{U }{2}(1- U)^2 \Phi^n_{i+ 1}
			+
		\frac{( 1 - U)}{2}  
		\left ( 3 -  (1 - U)^2 - 2 U^2
		\right ) \Phi^n_{i}
			\\ &\nonumber 
		+ \frac{U}{2} \left ( 3 - 2 (1 - U)^2 - U^2
			\right ) \Phi^n_{i- 1}
			-
		\frac{U^2}{2}  ( 1 - U)\Phi^n_{i- 2}
		\, .
	\end{align}
	The stability analysis 
	on \eqref{eq:cstUBEC} leads to the following amplification factor
	\begin{align}
		\xiB & = 1 - 2 i U \sin (\tfrac{1}{2}k \Delta x ) \Big[
		{\rm e}^{-\frac{1}{2} i k \Delta x}
		U (  1 
		+
		2 [ 1 - U ] \sin ^2 (\tfrac{1}{2}k \Delta x) )
		+
		\cos(\tfrac{1}{2}k \Delta x ) ( 1 - U )
		\Big]	 \, .
	\end{align}
	It can be shown analytically that the \BEC{} scheme is stable for $U\leq1$. 
	In fact, the \BEC{} 
	scheme is still stable for a CFL number smaller than $1.5$. 
	The truncation error analysis leads to
	\begin{align}
	  \label{eq:final_bec}
	    \sme{\Big[ \partial_t \Phi + u \partial_x\Phi \Big]}{\BEC} = 
	    	&
	    - (1- U)(1 - 2U ) 
	    \tfrac{u \Delta x^2}{3! } 
	    \partial^{3}_x \Phi
		 \\ &\nonumber
		    - 9 (1- U)^2 
	    \tfrac{u^2 \Delta x^2 \Delta t}{4! } \partial^{4}_x \Phi 
	  + ...
	\end{align}			

	Simulations with Heaviside, triangle and cosine distributions advected by a constant velocity were carried out for a CFL number $U>1$. For $U\lesssim 1.5$, the \BEC{} scheme gives finite results consistent with the stable results collected for $U<1$. The other schemes (\CIR , \FEC{} and \CEC) diverge for $U>1$ and the \BEC{} scheme diverges for $U \gtrsim1.5$. This extension of stability of the \BEC{} scheme can be understood in the following way: for $U>1$, the interpolation is performed with points that are not the nearest value to the reconstructed point. The contribution of the second nearest neighbors in the \BEC{} formula results in an enhanced stability of the scheme.

	The \FEC{} and \BEC{} schemes both have modified equations with a
	third order derivative truncation error. The \CEC{} scheme, 
	presented in \eqref{eq:intro:CEC} and \eqref{eq:intro:CECcoef}
	is a linear combination of these two schemes. The weights
	are computed to cancel the leading order of truncation error (see Appendix~A)
	and generate a higher order scheme.
	Using the linearity of the stability analysis, the amplification factor is
	\begin{align}
	\xiC = 1 
		&- \tfrac{2}{3} \sin(\tfrac{1}{2}k \Delta x)  \Big[
		U \Big( 3 + 2 [ 1 - U^2 ] \sin ^2 (\tfrac{1}{2}k \Delta x) \Big) 
		\sin (\tfrac{1}{2}k \Delta x ) 
			\\ & \nonumber
		+ \Big( 3 + 2 U [ 1 - U^2 ] \sin ^2 (\tfrac{1}{2}k \Delta x) \Big) 
		i \cos(\tfrac{1}{2}k \Delta x )
		\Big] \,.
	\end{align}	
    The \CEC{} scheme is stable for $U\leq 1$.
	To leading order, the modified equation of the \CEC{} scheme is
	\begin{align}
		\sme{\Big[  \partial_t \Phi 
		+ u \partial_x \Phi \Big]}{\CEC}
		\!\! =
		- (1 + U) ( 1 - U )(2 - U) 
		\tfrac{\vert u\vert (\Delta x)^3}{4! }
		\partial^4_x\Phi 
	+ ...
	\end{align}	

The essential properties of the different schemes
are reported in Tab.~\ref{tab:1dsumup}. The computational cost is evaluated 
using the number of composed interpolation operators.
The complexity of the interpolation operator varies 
with the interpolation method used. In the case of the \CIR{} scheme, 
the complexity is $\mathcal{O}(N)$ where $N$ is the total number grid of points.

\begin{table}[H]
	\centering
	\small
	\begin{tabular}{ c | c | c | c | c }
	\hline  \hline 
	Scheme & Formula 
	& Error & Stability & Nb of calls\\ 
	\hline

&&&& \\
\specialcell{ 
\CIR} & 
$\CIR 
\big[\Phi \big] 
\!=\! L_{+} \big[\Phi \big]$ & 
$\left( 1- U \right) \frac{ \vert u\vert \Delta x }{ 2} \partial^2_x \Phi $ & 
$U<1$ &
$1$ \\ 
&&&& \\

\FEC & 
$\FEC 
\big[\Phi \big] \!\!=\!\! 
L_{+}\big[ \Phi \big] + \tfrac{1}{2} (\Phi-\bar{\Phi})$ & 
\specialcell{$
- ( 1- U^2 ) \frac{u \Delta x^2}{3!} \partial^{3}_x \Phi$ \\[2mm]
$- 3(1- U^2) \frac{u^2 \Delta x^2 \Delta t }{4!} \partial^{4}_x \Phi $ }&
$U<1$ &
$2$ \\
&&&& \\

\BEC & 
$\BEC 
\big[\Phi \big] \!\!=\!\!
L_{+} \big[\Phi + \tfrac{1}{2} (\Phi-\bar{\Phi}) \big]$ & 
\specialcell{$- (1- U)(1 - 2U ) \tfrac{u \Delta x^2}{3! } \partial^{3}_x \Phi $ \\[2mm]
$- 9 (1- U)^2 \tfrac{u^2 \Delta x^2 \Delta t}{4! } \partial^{4}_x \Phi $} & 
$U\lesssim 1.5$ &
$3$ \\
&&&& \\

\CEC & 
\specialcell{ 
$\CEC 
\! \big[\Phi \big] \!\! = \!\!
L_{+} \!\Big[ \! 
\Phi +\tfrac{1+U \!\! }{6U \!\! } 
(\Phi-\bar{\Phi} \!) \!\Big] $ \\[2mm]
$+ \tfrac{1-2U}{6U} (\Phi-\bar{\Phi})$} & 
$ - (1 + U) ( 1 - U )(2 - U) 
\tfrac{\vert u \vert (\Delta x)^3}{4! }
\partial^4_x\Phi $ &
$U < 1$ &
$3$ \\[7mm]
\hline
\end{tabular} 
	\caption{Comparative table of one dimension schemes.}
	\label{tab:1dsumup}
\end{table}

\section{Results for one-dimensional problems}
\label{sec:res1D}
\begin{figure}[H]
	\centering
	\subfigure[$10$ periods]{
	\includegraphics[scale=0.36,trim= 20 10 45 40, clip=true]
	{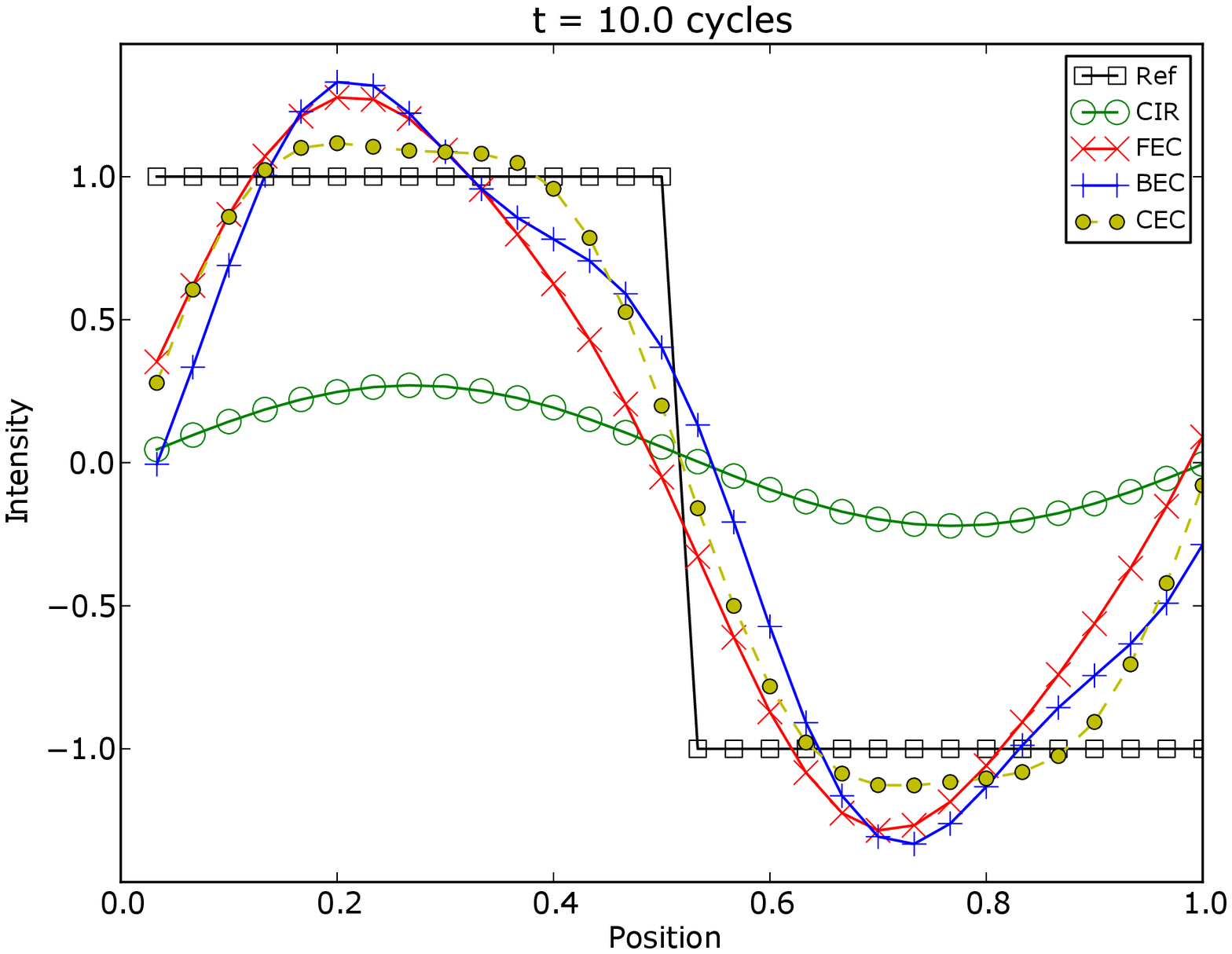}
	}
	\subfigure[$100$ periods]{
	\includegraphics[scale=0.36,trim= 20 10 45 40, clip=true]
	{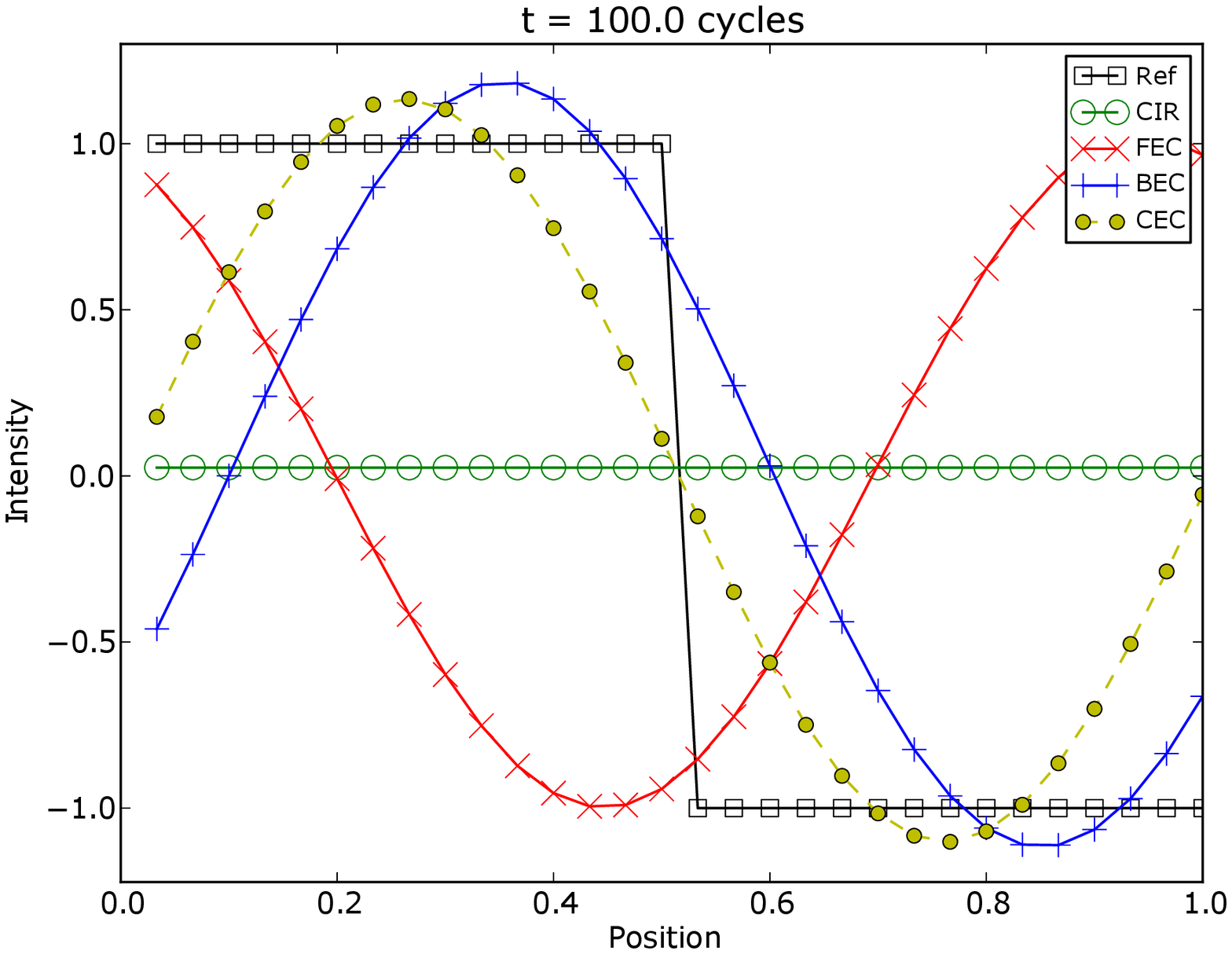}
	}
	\caption{One dimension advection of a Heaviside with a resolution of $N=30$ at $CFL=0.75$.}
	\label{fig:1Dhea}
\end{figure}

\begin{figure}[H]
	\centering
	\subfigure[$10$ periods]{
	\includegraphics[scale=0.36,trim= 20 10 45 40, clip=true]
	{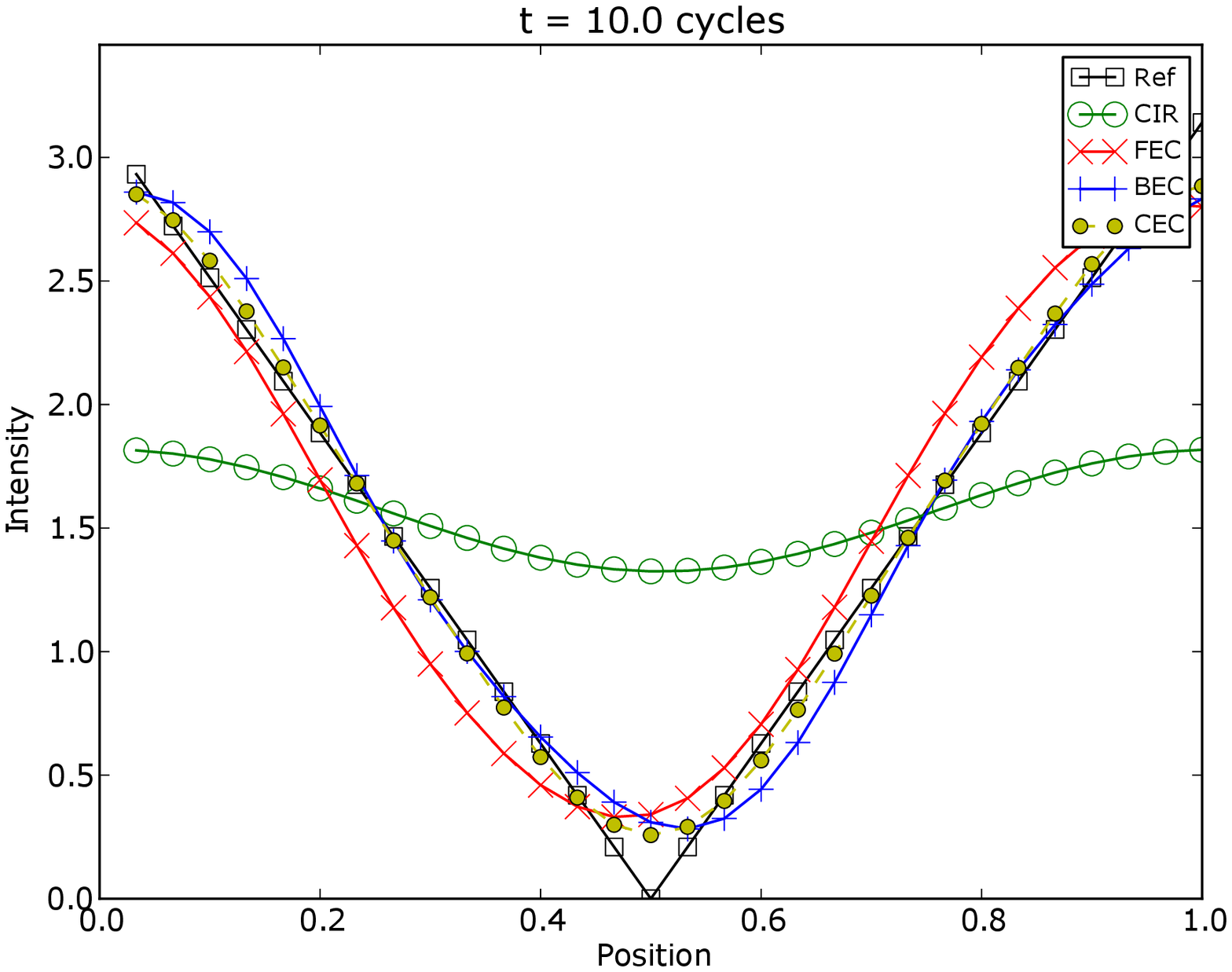}
	}
	\subfigure[$100$ periods]{
	\includegraphics[scale=0.36,trim= 20 10 45 40, clip=true]
	{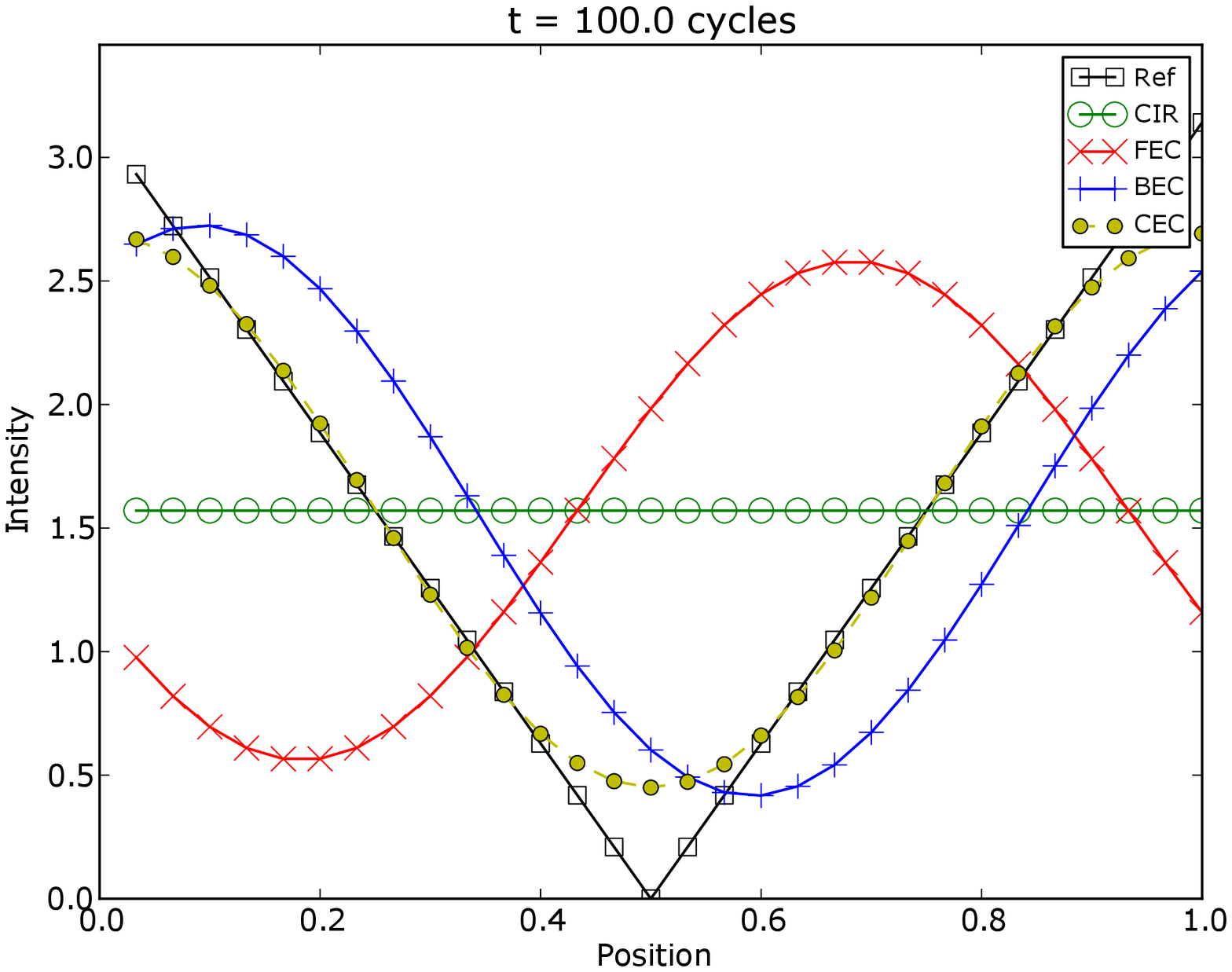}
	}
	\caption{One dimension advection of a triangle with a resolution of $N=30$ at $CFL=0.75$.}
	\label{fig:1Dtri}
\end{figure}

\begin{figure}[H]
	\centering
	\subfigure[$10$ periods, i.e. $1200$ time-steps]{
	\includegraphics[scale=0.36,trim= 20 10 45 40, clip=true]
	{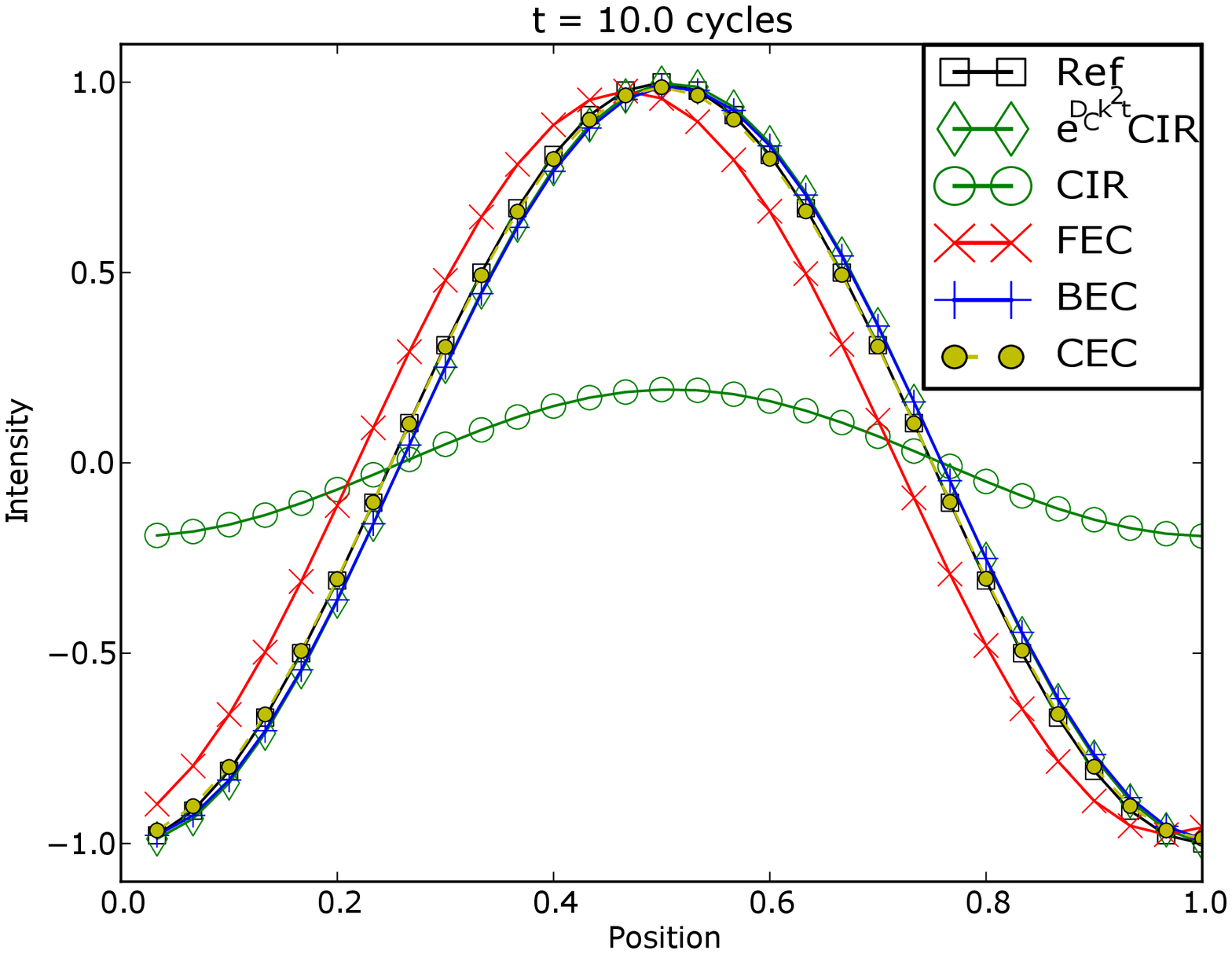}
	}
	\subfigure[$100$ periods, i.e. $12000$ time-steps]{
	\includegraphics[scale=0.36,trim= 20 10 45 40, clip=true]
	{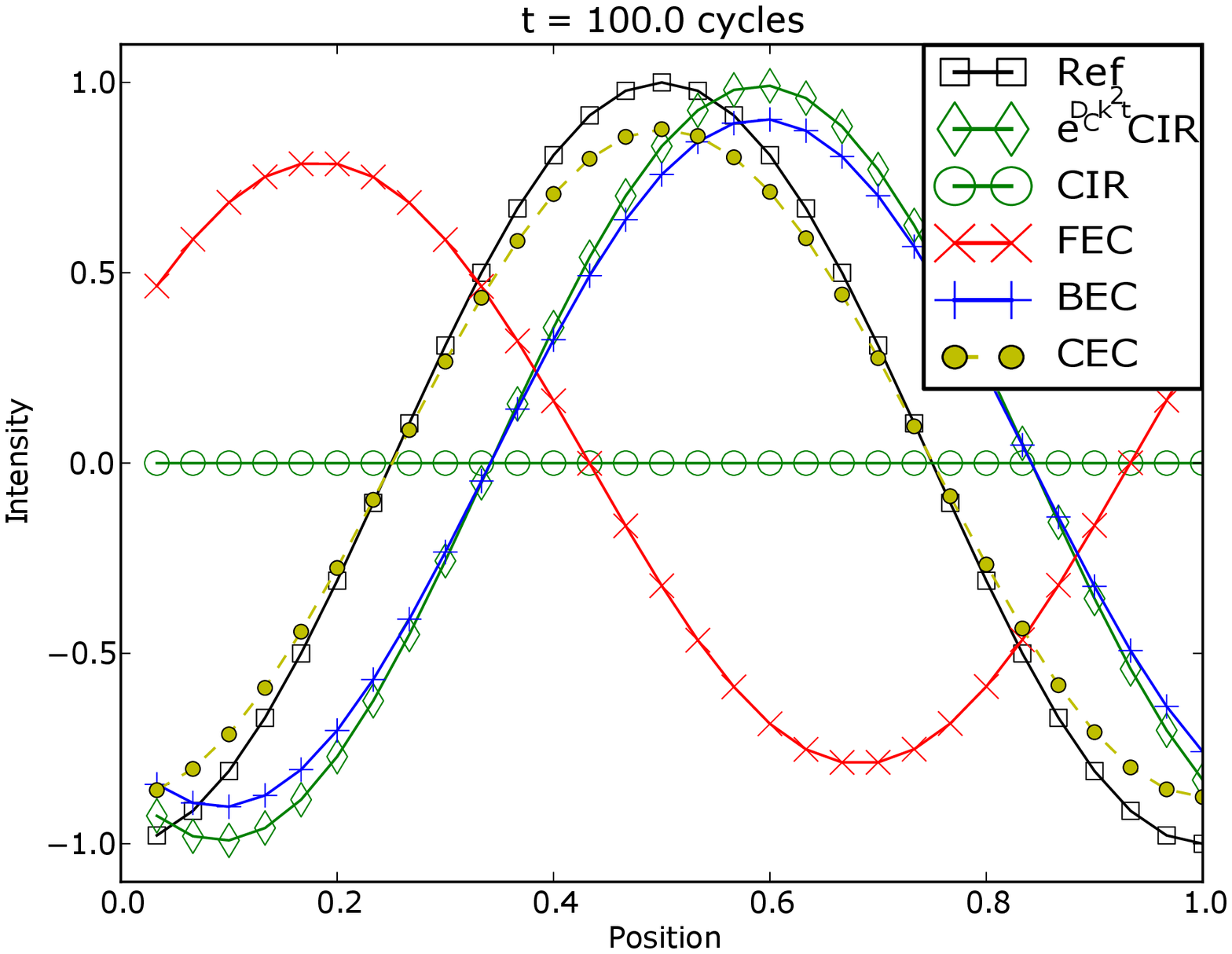}
	}
	\caption{One dimension advection of the cosine function with a resolution of $N=30$ at $CFL=0.75$.}
	\label{fig:1Dcos}
\end{figure}

	To assess the efficiency of the schemes introduced previously, simulations
	with a constant velocity were performed. A one-dimensional periodic domain is considered, and the solution is advected for 10 or 100 cycles. Fig.~\ref{fig:1Dhea}, \ref{fig:1Dtri} and \ref{fig:1Dcos} 
	show the advection of three density profiles with different regularities. 
	Because of the Fourier properties of sine functions, the first harmonic was
	studied thoroughly to check that it matches the properties of the modified equation. 

	The first set of tests was performed using an Heaviside profile
	$
	\Phi(x,t=0) = \sgn \Big[ \sin \left (
	2\pi x / l \right ) \Big] \, .
	\label{eq:hea}
	$
        This is a demanding test, as this profile
	is discontinuous at two cross-over positions ($0$ and $0.5$). 
	As time elapses, the high frequencies get damped 
	and the profile is nearly reduced to its first harmonic. 
	In fig.~\ref{fig:1Dhea},
	the \CEC{} scheme is closer to the 
	analytical solution than the other schemes by three criteria: (i) 
	the ``flatness'' of its profile at the beginning of the simulation, (ii) 
	the distance from the analytic cross-over position at all time and
	(iii) the phase drift of the profile at long time. These criteria may seem independent but they 
	are all linked to the Fourier properties of the modified equation.

	The second set of tests was performed using a triangular profile, 
		$
	\Phi(x,t=0) = \vert x / l - 0.5 \vert \, .
	\label{eq:tri}
		$
	This profile
	is non differentiable at two cross-over position ($0$ and $0.5$). 
	In fig.~\ref{fig:1Dtri}, the observations reported in 
	the previous paragraph still hold for the triangular profile. As expected,
	the \CEC{} scheme is closer to the analytic results 
	in the case of a continuous but non-derivable profile.

	The last tests were performed using the first harmonic cosine profile,
		$
	\Phi(x,t=0) = - \cos \left ( 2\pi {x}/{l} \right ) \,.
	\label{eq:cos}
		$
	The properties of 
	the profile will be studied in more details in 
	fig.~\ref{fig:1DresplotDR} and \ref{fig:1DresplotPD}. 
	In fig.~\ref{fig:1Dcos}, the \CIR{} scheme is
	so diffusive that a ``corrected \CIR{}'' (green diamond
	line)\footnote{The corrected \CIR{} values are
	equal to those of \CIR{} multiplied by ${\rm exp}({ \sme{D}{\CIR} \, k^2 t })$ 
	where $\sme{D}{\CIR}$ is defined in \eqref{eq:CIRdiff}.} is plotted.
	Even though the \CIR{} scheme is near zero in fig.~\ref{fig:1Dcos},
	the norm of its difference to the analytic profile is smaller than the \FEC{} scheme
	which drifted to such an extent that it is nearly opposite to the
        reference profile.

\begin{figure}[H]
	\centering
	\subfigure[$10$ periods, i.e. $80$ time-steps ]{
	\includegraphics[scale=0.36,trim= 20 10 45 40, clip=true]
	{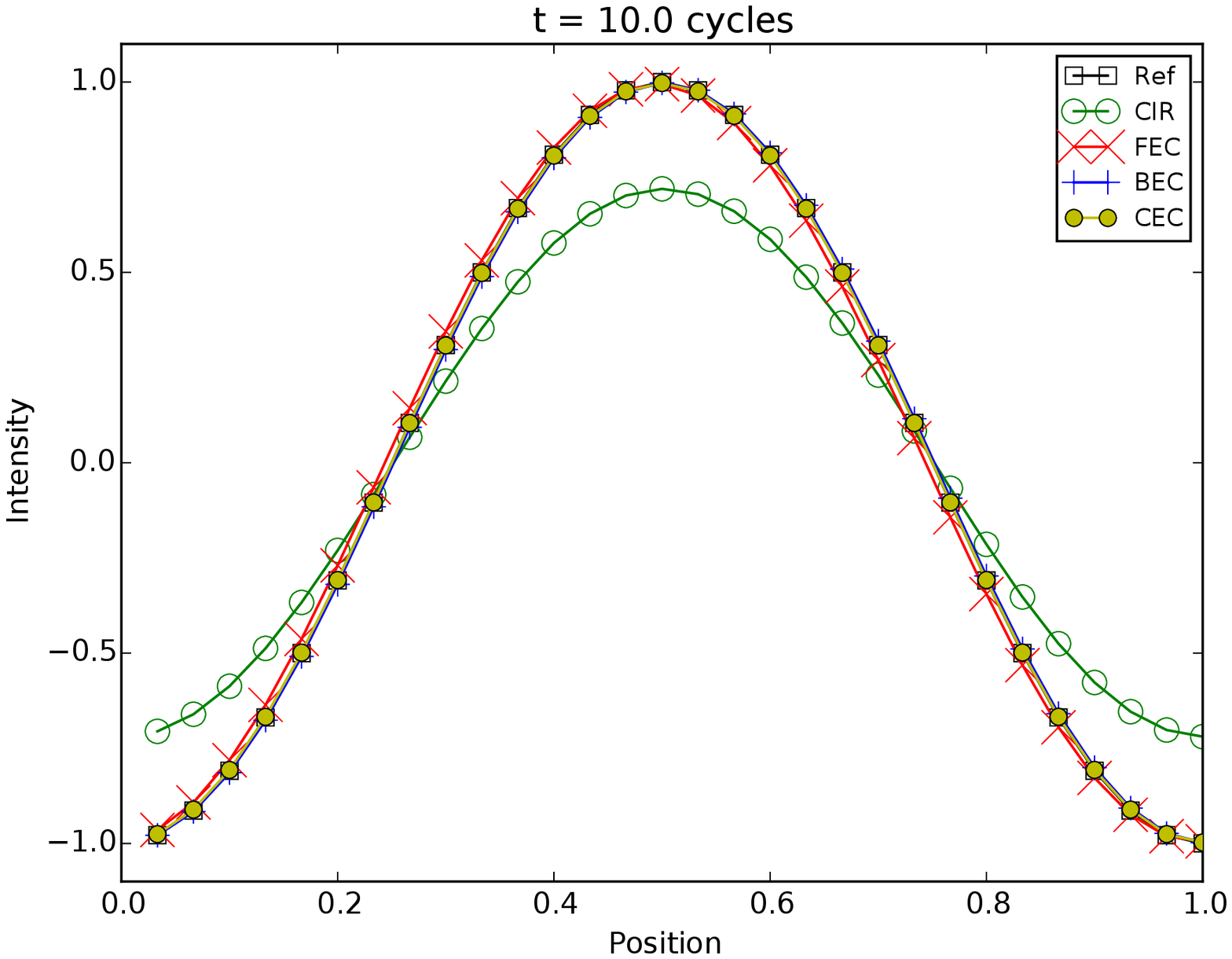}
	}
	\subfigure[$100$ periods, i.e. $800$ time-steps ]{
	\includegraphics[scale=0.36,trim= 20 10 45 40, clip=true]
	{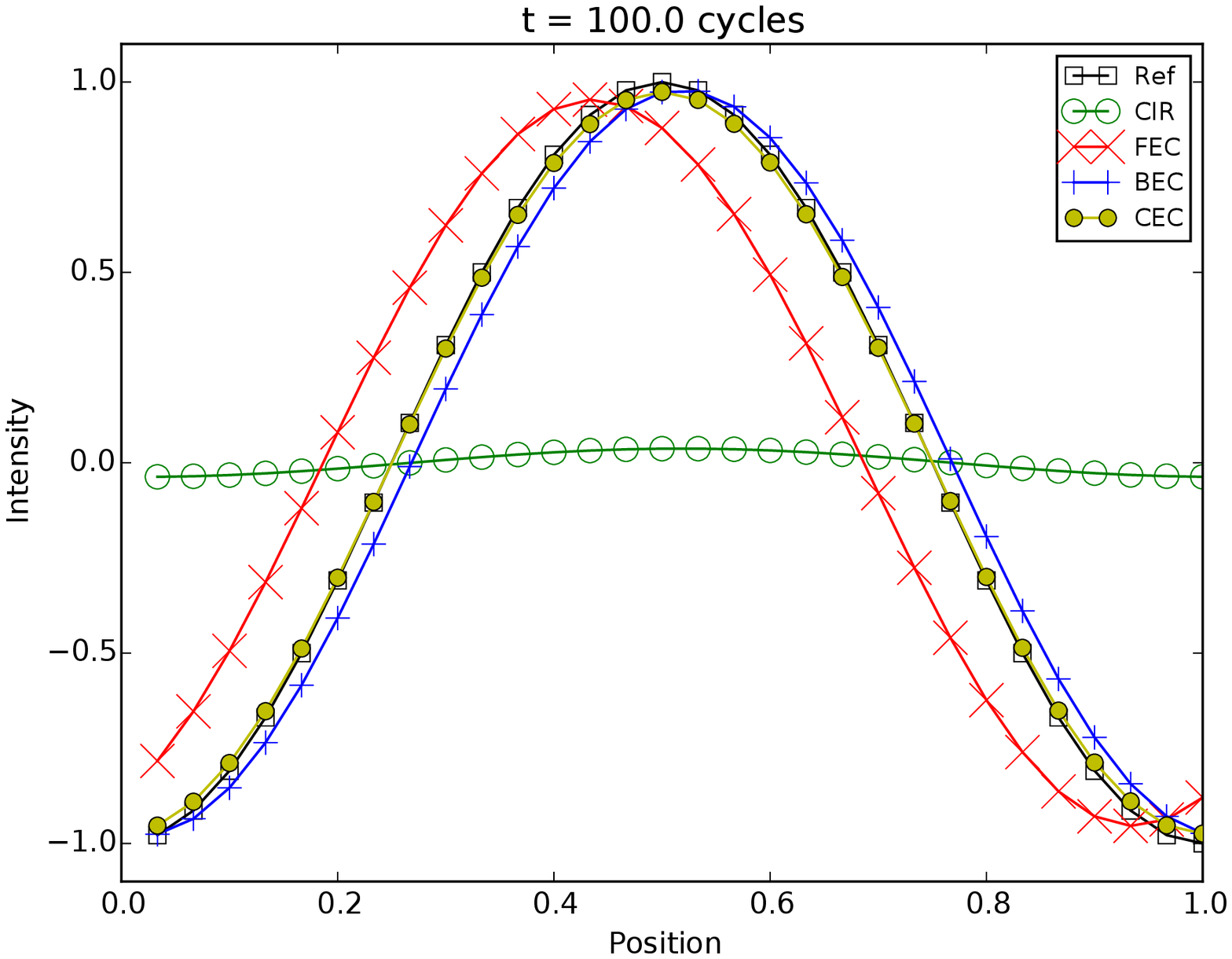}
	}
	\caption{One dimension advection of the cosine function with a resolution of $N=30$
  at $CFL=3.75$.}
	\label{fig:1Dcfl}
\end{figure}

        As noted above, provided the interpolation strategy involves
        non-neighboring points, semi-Lagragian methods can use $CFL$ number
        larger than one. Using a non-local interpolation stencil, we can
        reproduce the advection test of fig.~\ref{fig:1Dcos} using a $CFL$
        number of
        $3.75$, see fig.~\ref{fig:1Dcfl}.

        The time-step being larger in this last case, fewer time-steps are needed 
        for the same integration time (here respectively $10$ and $100$ periods),
        the effects of numerical dispersion and diffusion
        are thus weakened compared to fig.~\ref{fig:1Dcos}

        This is achieved with a simple  modification of relations
        \eqref{eq:intro:CECcoef} to compute the weights $\cA{}$ and $\cB{}$,
        in the form
	\begin{align}
		3 \, \cA = 2 - \frac{1}{\left( \vert u \vert \Delta
                  t/\Delta x \right) \, \% 1}
		\quad \text{and} \quad
		\cB =1 - \cA 
		\, , 
      \label{eq:intro:CECcoefSL}
	\end{align}
        (where $\%1$ denotes the remainder of the
        division by unity),
        the accuracy of the \CEC{} scheme is preserved for large CFL numbers.

\section{Multi-dimensional problems}
\label{sec:res2D}
The extension of the above procedures to multi-dimensional problems requires some care.
For instance in two dimensions, the \CIR{} scheme is
	\begin{align}
		\label{eq:CIR2D}
		\CIR[\Phi]_{i,j} 
		= &
		( 1 - U_{i,j}^x )(1 - U_{i,j}^y)\Phi^n_{i,j}
		+ 
		(1 - U_{i,j}^x )U_{i,j}^y \Phi^n_{i,j - s_{i,j}^y} 
			\, ,\\ \nonumber	
		&+
		U_{i,j}^x( 1 - U_{i,j}^y )\Phi^n_{i- s_{i,j}^x , j} 
		+
		U_{i,j}^x U_{i,j}^y\Phi^n_{ i - s_{i,j}^x , j - s_{i,j}^y }
		\, .
	\end{align}
The semi-Lagrangian \CIR{} scheme uses one more value ($\Phi[ i - s_{i,j}^x ][ j - s_{i,j}^y ] $) than the Eulerian Upwind scheme. 
However, the \CIR{} scheme is very similar to the 
directionally split Upwind scheme
	\begin{align}
		\Phi^{\star}_{i,j}= 
		&
		( 1 - U_{i,j}^x) \Phi^n_{i,j} + U_{i,j}^x 
		\Phi^n[i- s_{i,j}^x][j] 
			\, , 		
    \label{eq:split:1} \\
		\Phi^{\star\star}_{i,j} = 
		&
		( 1 - U_{i,j}^y) \Phi^{\star}_{i,j}
		+ U_{i,j}^y 
		\Phi^{\star}[i][j- s_{i,j}^y] \, .
		\label{eq:split:2}
	\end{align}
In the general case in multi-dimension, there is no expression for the $\cA$ and $\cB$ coefficients of the \CEC{} scheme.
It can be extended to any dimension if the scheme is directionnally split as done in \eqref{eq:split:1} and \eqref{eq:split:2}. However, if a simple
splitting method is used, the approximation is reduced to first
order. Special splitting methods, such as Strang splitting 
\cite{strang_construction_1968}, are required to increase the order of the total scheme.

\label{sec:2dres:rotdif}
\begin{figure}[H]
	\centering	
	\subfigure[Initial patch distribution]{
	\includegraphics[height=\figcma, trim= 20 10 60 10, clip=true]
	{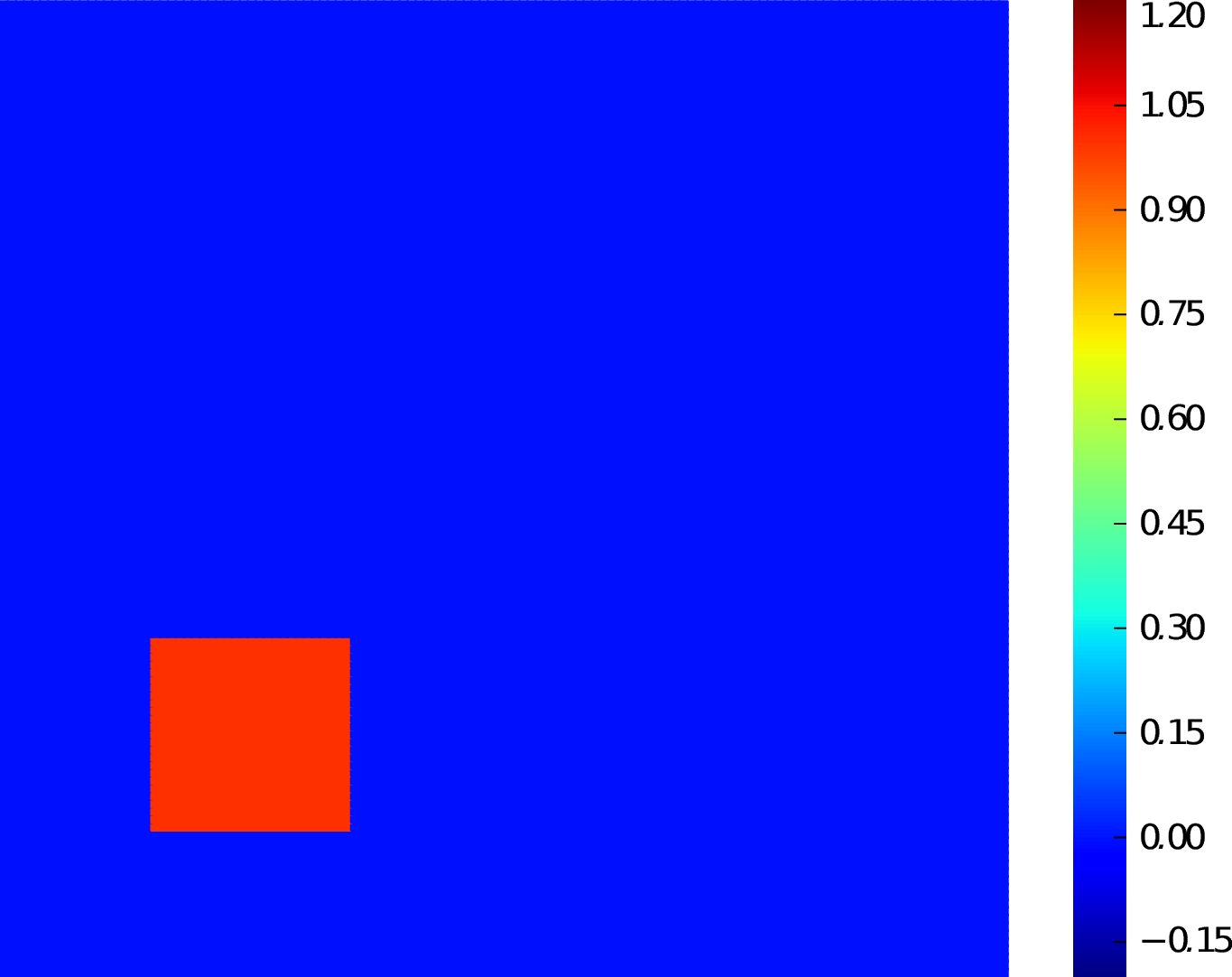} 
	\label{fig:inipatch} }
	\subfigure[Velocity profile $u(0,y)$ or $v(x,0)$]{
	\includegraphics[height=\figcma, trim= 20 10 10 10, clip=true]
	{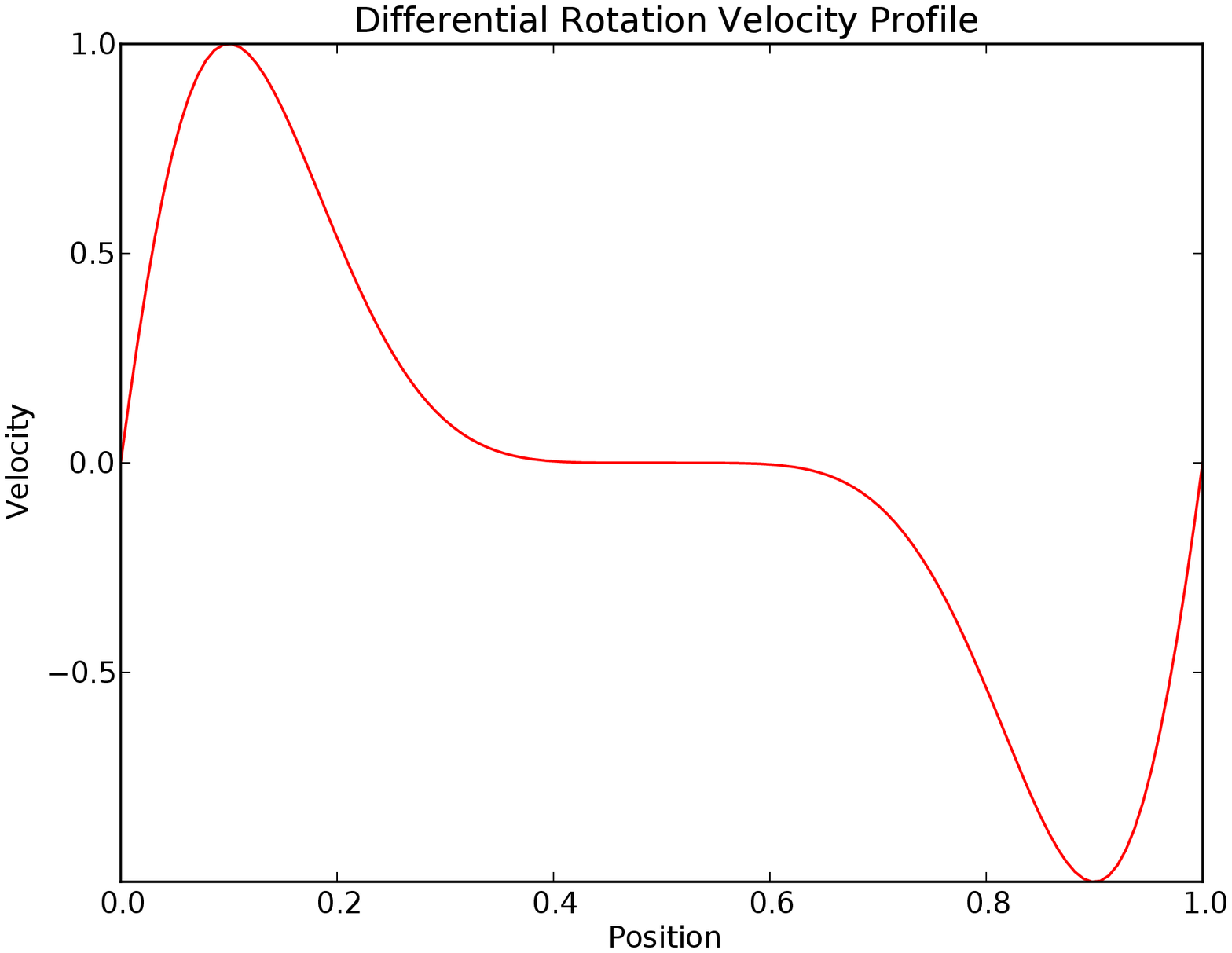}
	\label{fig:velgraph} }
	\subfigure[Pure Lagrangian advection]{
	\includegraphics[height=\figcma]
	{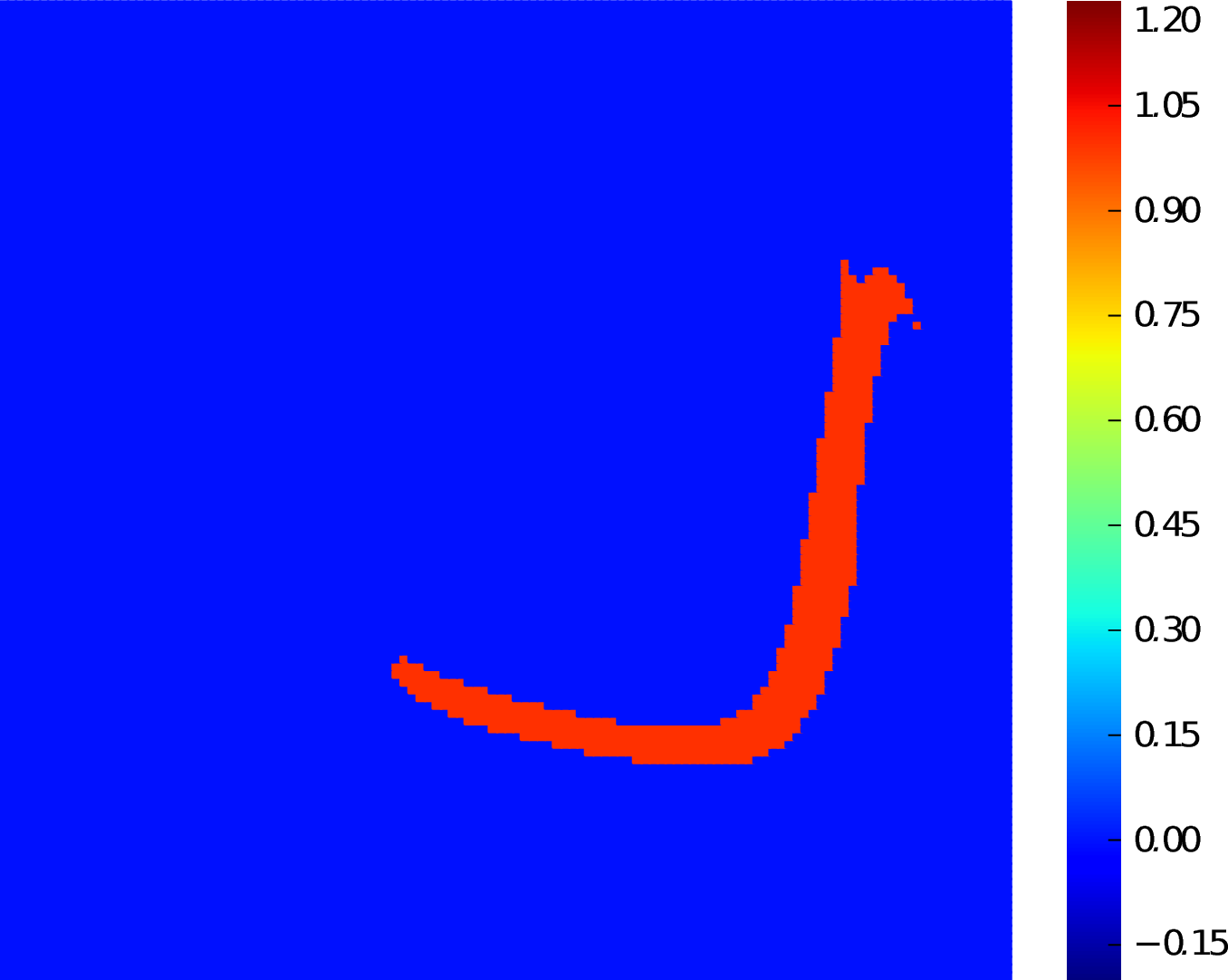} 
	\label{fig:endpatch} }
	\caption{Initial condition, velocity profile and final distribution for the two-dimensional advection test case.}
\end{figure}
To illustrate applications of our strategy to higher dimensions, let us consider
an advection problem in two dimensions of space. 
A squared patch is considered for the initial distribution of the passive scalar: one inside the square 
and zero outside, as presented in fig.~\ref{fig:inipatch}. 
The order of the schemes for regularly varying 
velocities should be the same as the one for constant velocities. 
Quantitative results being difficult, only qualitative observations will be made.
The following velocity field was used to test the schemes 
\begin{align}
	u(x,y) = \quad& 
	\frac{y}{l} \left(1- \frac{y}{l} \right) \left(\frac{1}{2}- \frac{y}{l}\right) 
	\left[\cos \Big( 2\pi \frac{y}{l} (1-\frac{y}{l})  \Big)+1 \right]/(2 \pi^2)
	\, ,
	\\
	v(x,y) = - & 
	\frac{x}{l} \left(1-\frac{x}{l}\right) \left(\frac{1}{2}-\frac{x}{l}\right) 
	\left[\cos\Big( 2\pi \frac{x}{l}(1-\frac{x}{l}) \Big) + 1 \right]/(2 \pi^2)
	\, ,
\end{align}
where $l$ is the length of the box in both directions.
In fig.~\ref{fig:velgraph}, the velocity cancels out on 
the edges of the box and is divergence free. 
With the profiles used, the patch
is not transported through the walls of the box even though the simulation has
periodic boundary conditions. The patch never intersects itself which 
makes it easier to track. To compare the results, 
a fully Lagrangian method was used as a reference. 
The time-step of this method was twenty times smaller to have more accurate results.
The solution is represented in fig.~\ref{fig:endpatch}.

\begin{figure}[H]
	\centering
	\subfigure[\CIR{} advection]{
	\includegraphics[width=\figcm, height=\figcm, 
		trim=80 50 90 80, clip=true, keepaspectratio=false]
                {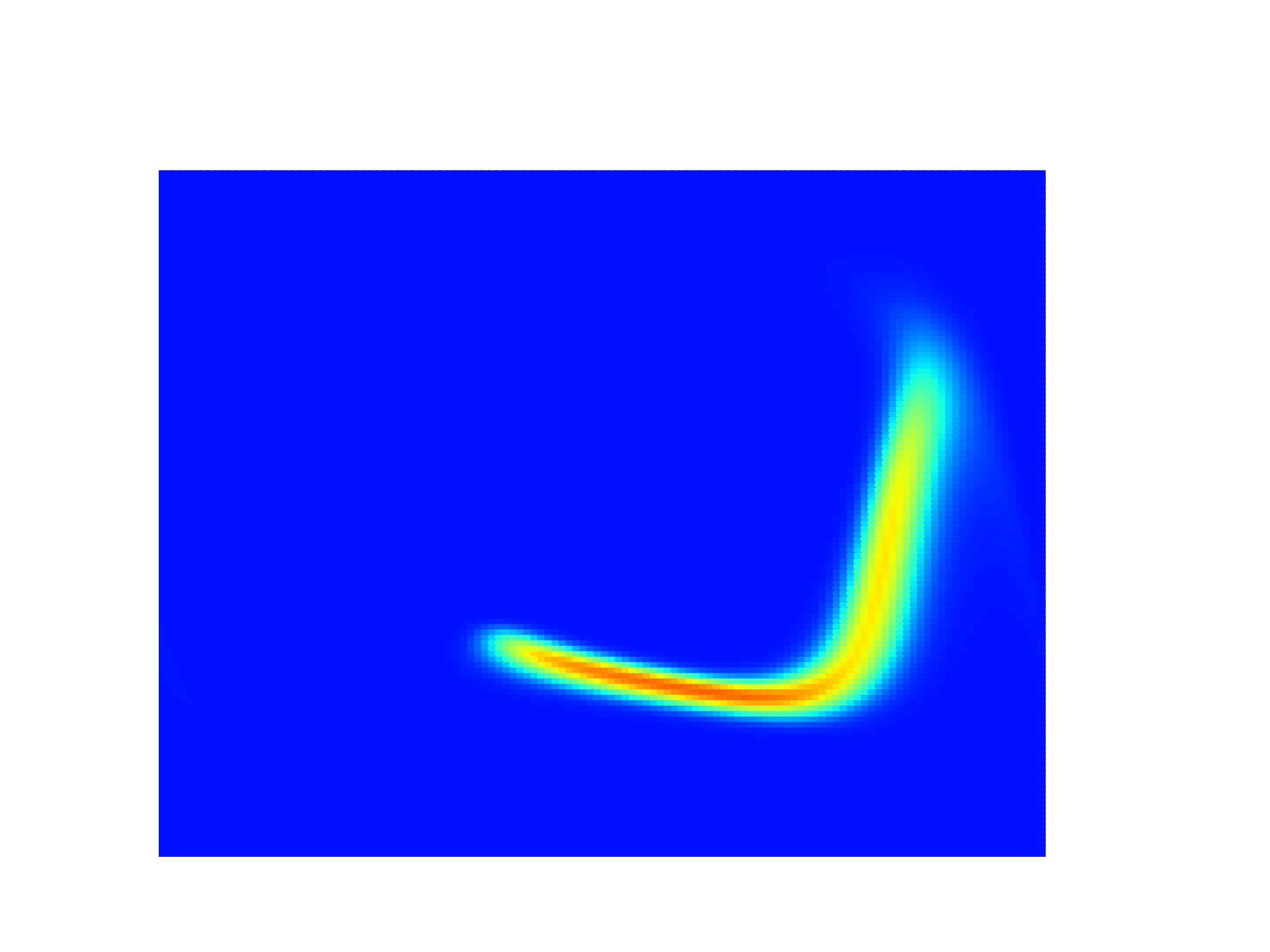} }	
    \subfigure[\FEC{} advection]{
	\includegraphics[width=\figcm, height=\figcm, 
		trim=80 50 90 80, clip=true, keepaspectratio=false]
                {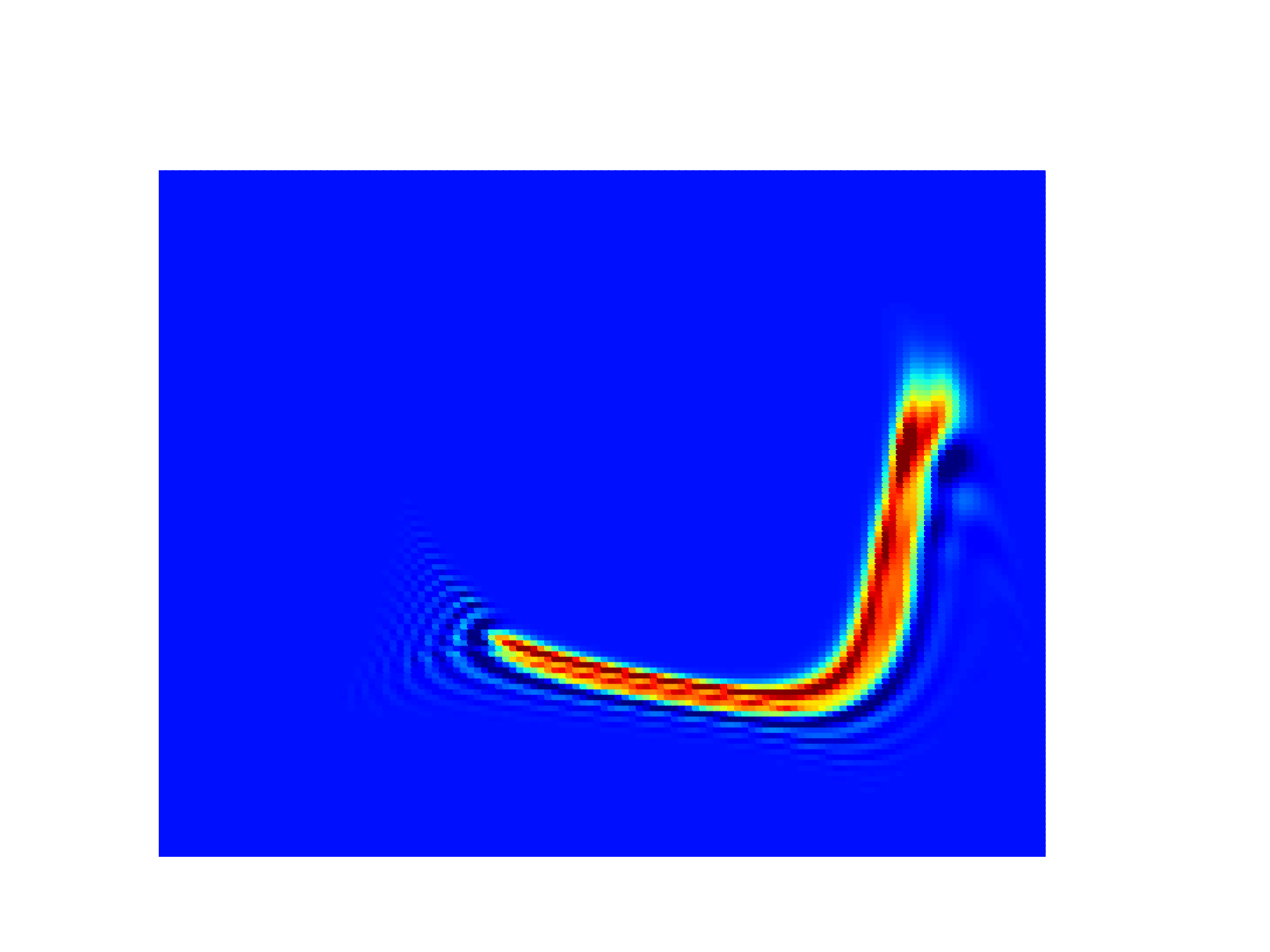} }
	\subfigure[\BEC{} advection]{
	\includegraphics[width=\figcm, height=\figcm, 
		trim=80 50 90 80, clip=true, keepaspectratio=false]
                {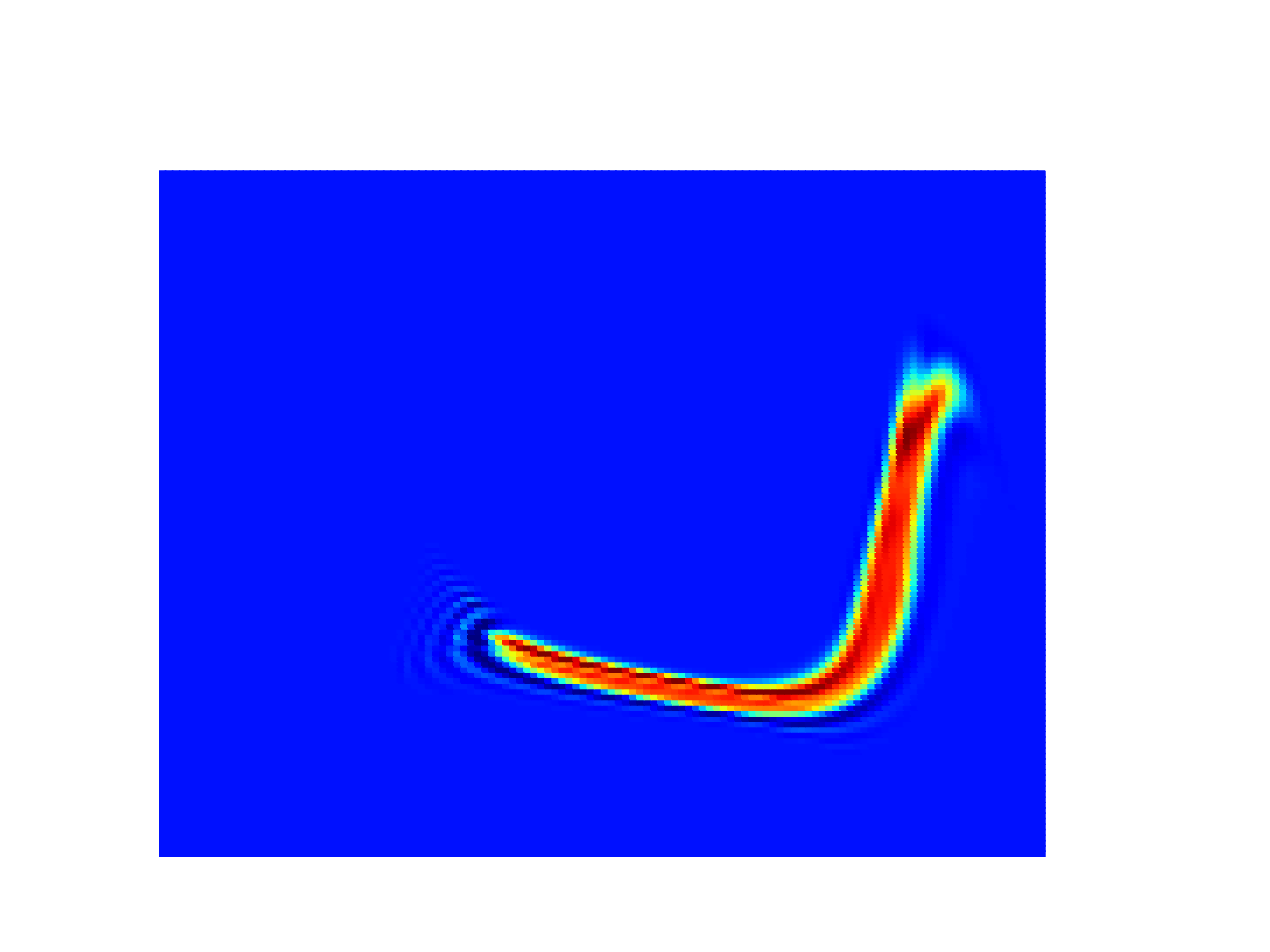} }
	\subfigure[\CEC{} advection]{
	\includegraphics[width=\figcm, height=\figcm, 
		trim=80 50 90 80, clip=true, keepaspectratio=false]
				{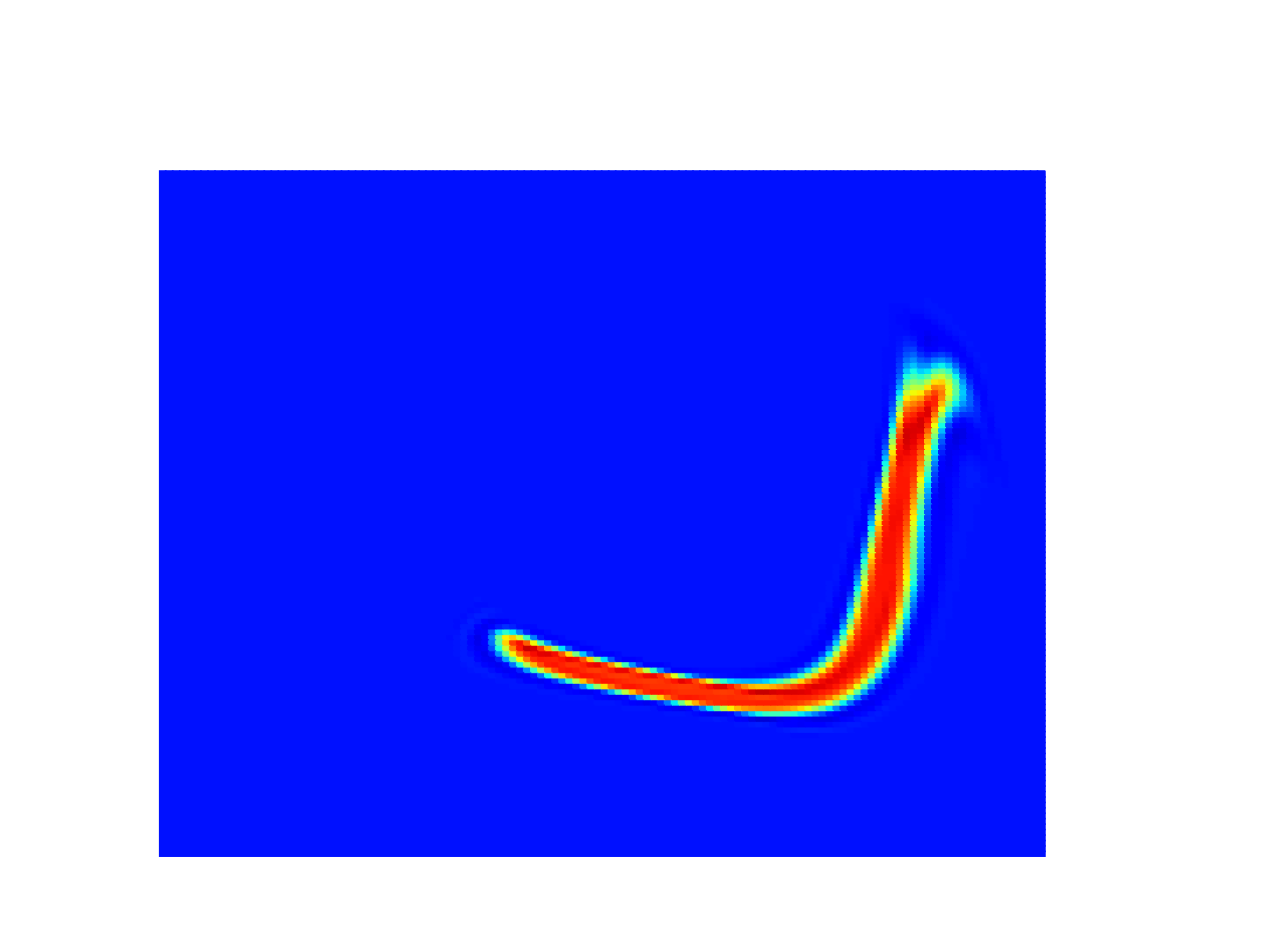} }
	\caption{Two-dimensional patch advection using the different schemes.}
    \label{fig:dis1}
\end{figure}

\begin{figure}[H]
	\centering
	\subfigure[\CIR{} error.]{
	\includegraphics[width=\figcm, height=\figcm, 
		trim=80 50 90 80, clip=true, keepaspectratio=false]
                {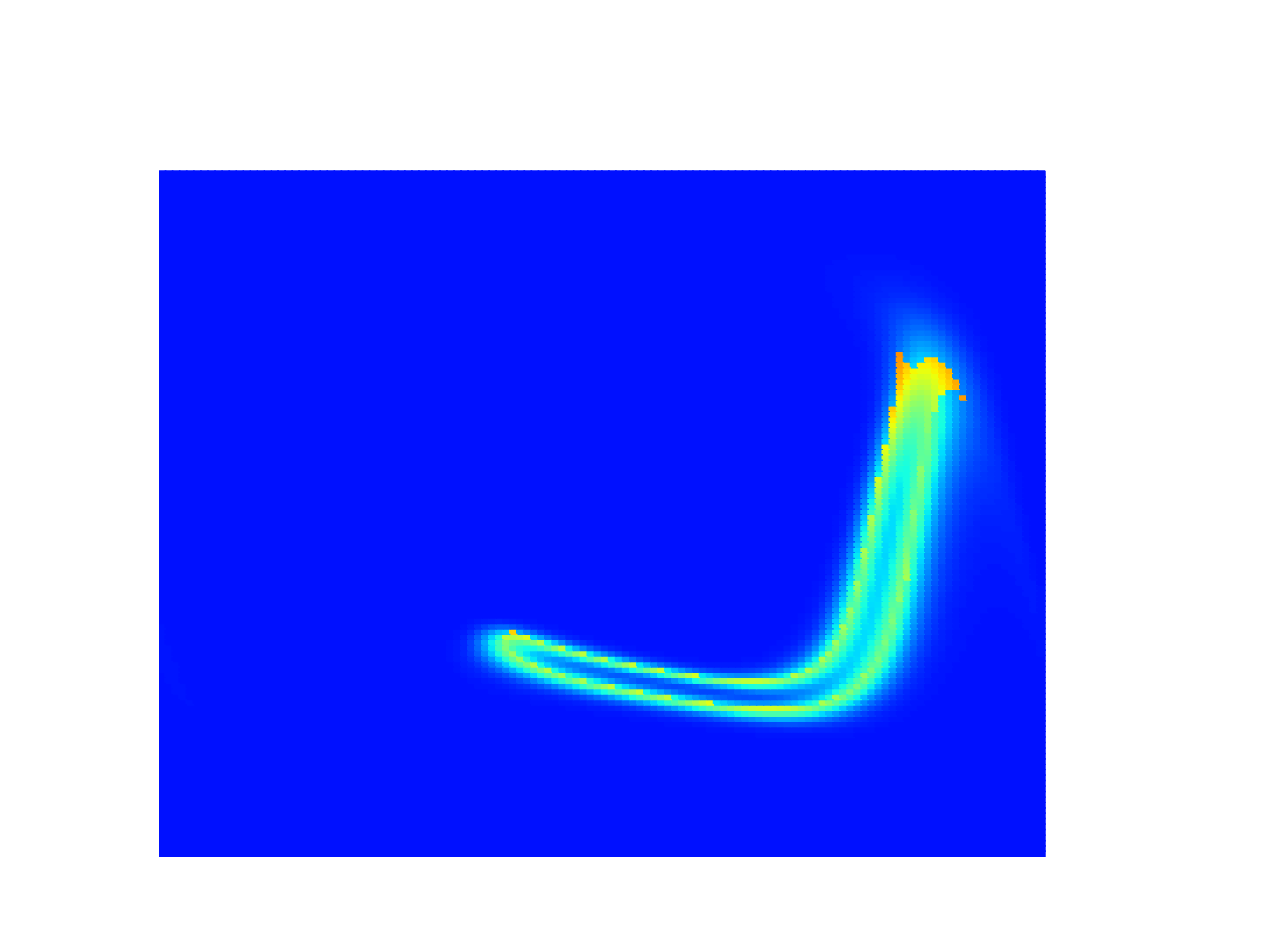}
    \label{fig:disCIR}}
	\subfigure[\FEC{} error.]{
	\includegraphics[width=\figcm, height=\figcm, 
		trim=80 50 90 80, clip=true, keepaspectratio=false]
                {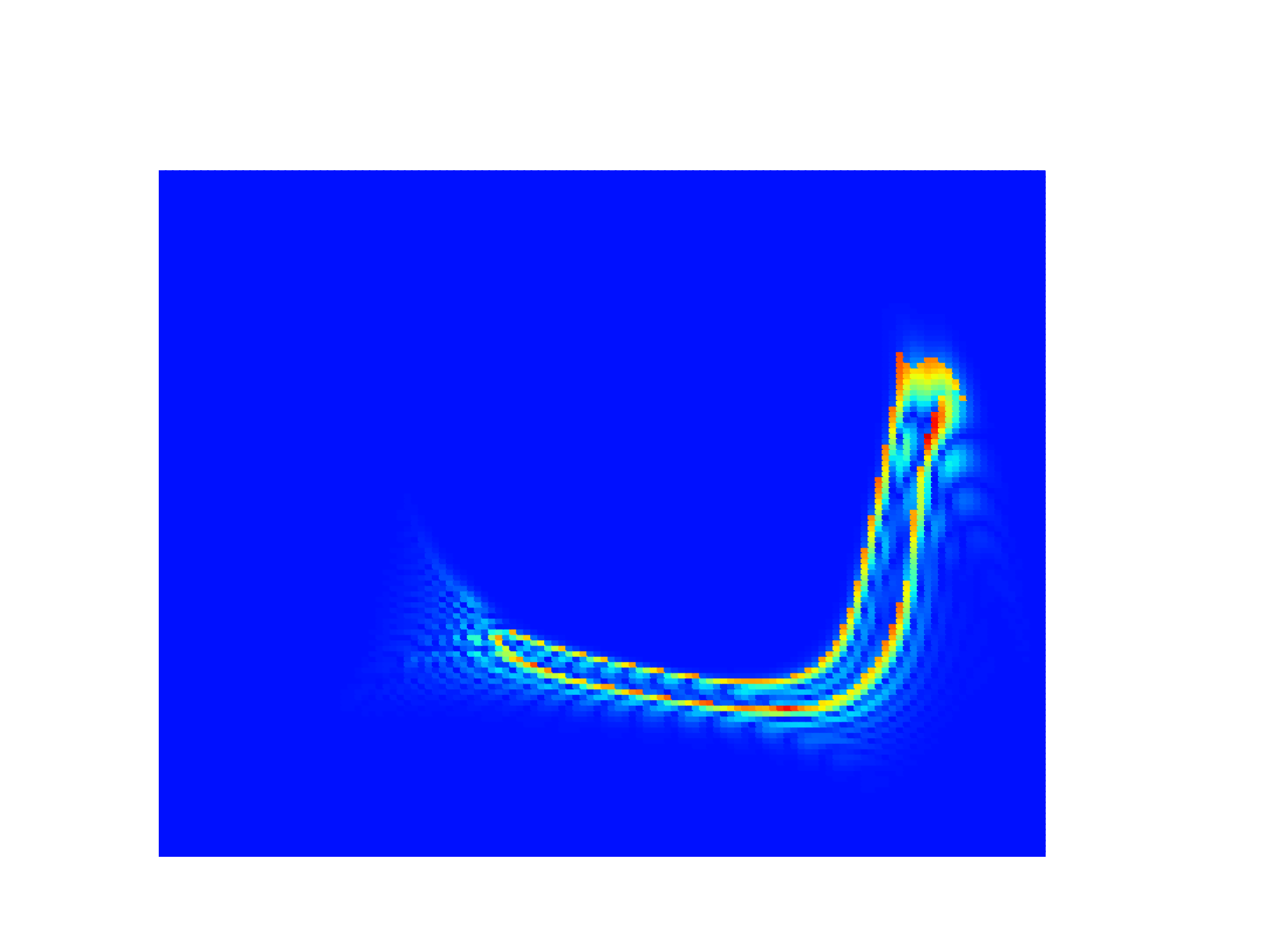} 
    \label{fig:disFEC}}
	\subfigure[\BEC{} error.]{
	\includegraphics[width=\figcm, height=\figcm, 
		trim=80 50 90 80, clip=true, keepaspectratio=false]
                {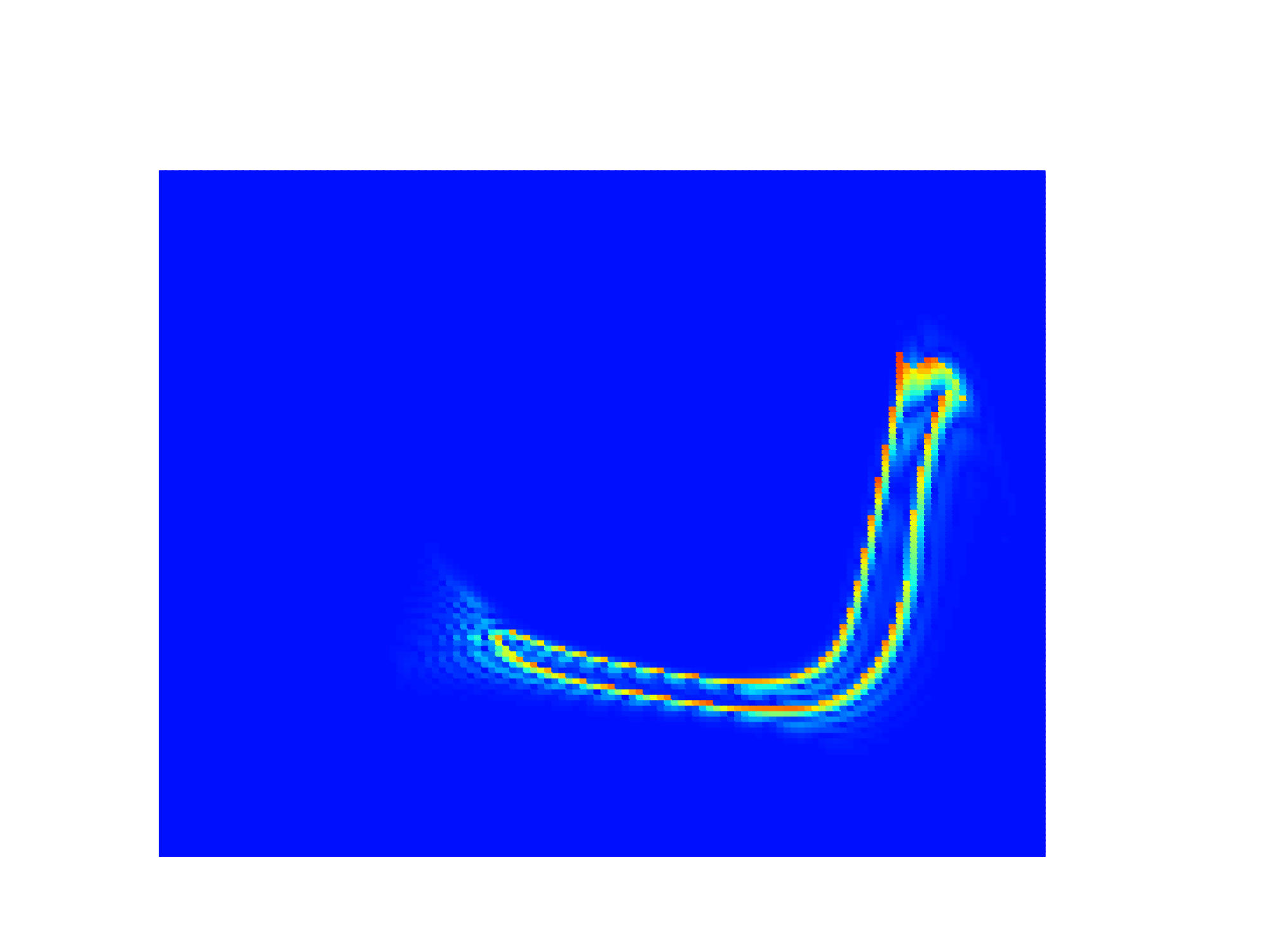} 
    \label{fig:disBEC}}
	\subfigure[\CEC{} error.]{
	\includegraphics[width=\figcm, height=\figcm, 
		trim=80 50 90 80, clip=true, keepaspectratio=false]
		{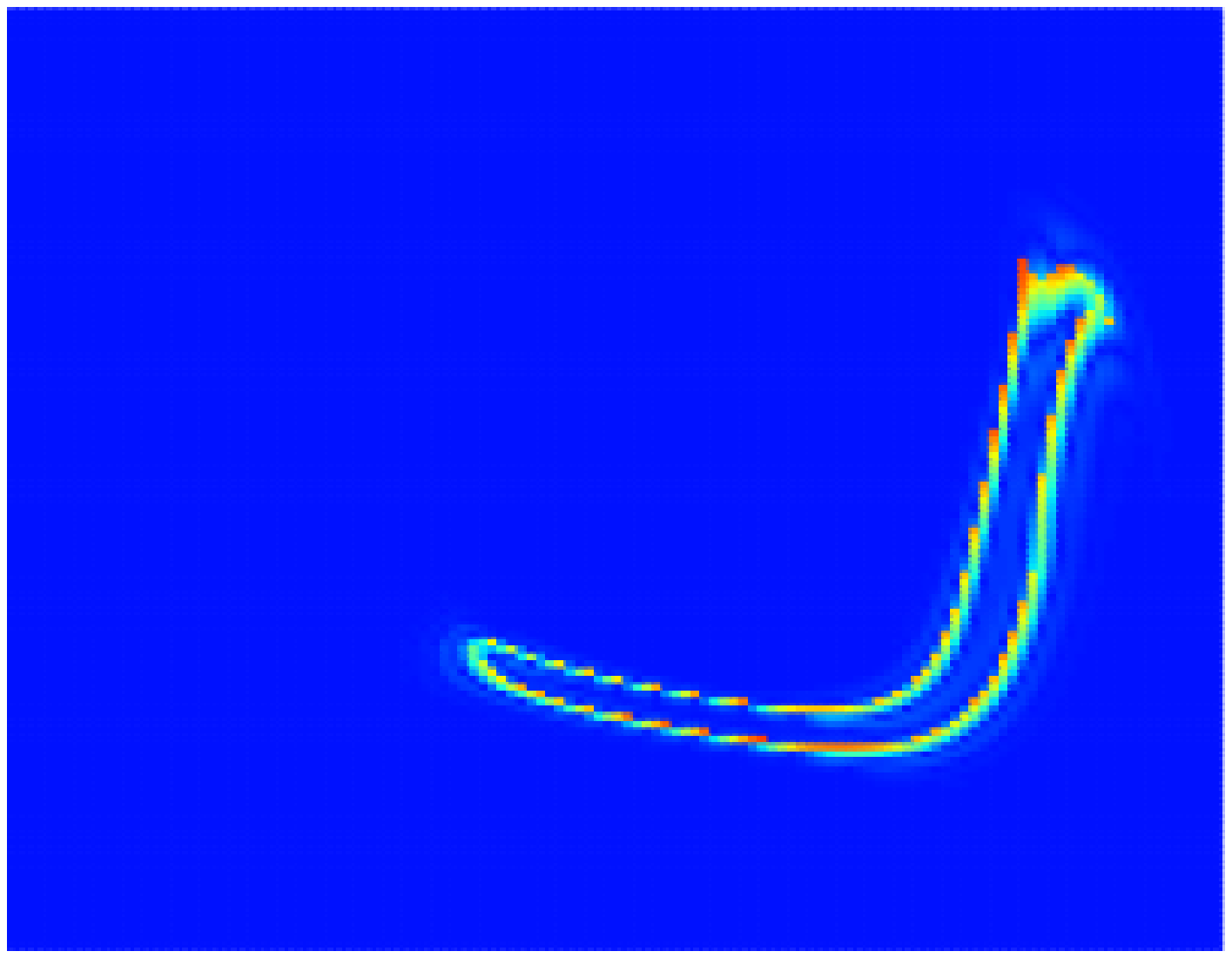} 
    \label{fig:disCEC} }
  	\caption{Error, as measured by the difference of the numerical
          solutions to the reference solution obtained with pure Lagrangian advection.}
    \label{fig:dis2}
\end{figure}

In fig.~\ref{fig:dis1} and \ref{fig:dis2}, the analysis of the gap between a scheme and 
the reference solution should not only be guided by the intensity of the 
difference but also by the area impacted. 
The \CIR{} scheme clearly introduces the largest computational error.

The perturbation of the distribution can also give an intuition of the leading error term
in the modified equation. The quick oscillations at the tail of the patch in fig.~\ref{fig:disFEC} 
and \ref{fig:disBEC} can be related to the dispersive residuals of 
the \FEC{} and \BEC{} schemes.
In fig.~\ref{fig:disCEC}, the \CEC{} solution is the closest to the reference solution obtained by the pure Lagrangian method. 
The error is of small amplitude and only impacts the edges of the patch. 

\section{Application to thermal convection}
\label{sec:therConv}

\begin{figure}[H]
	\centering
	\subfigure[\CIR{} scheme]{
	\includegraphics[scale=0.5, trim=70 30 160 30, clip=true]
                  {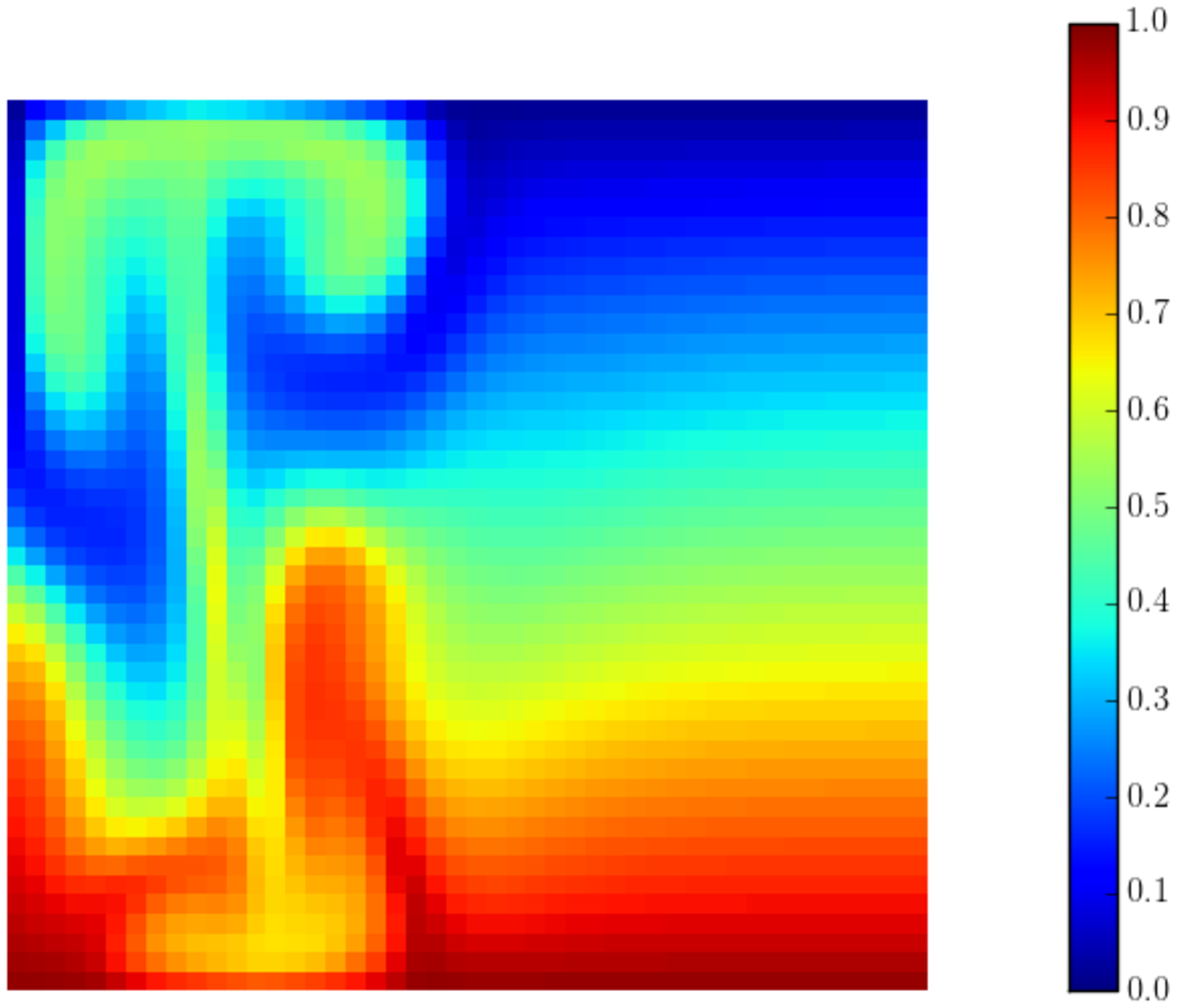}
                  \label{sf:CIR50}
	}
	\subfigure[\FEC{} scheme]{
	\includegraphics[scale=0.5, trim=70 30 80 30, clip=true]
                  {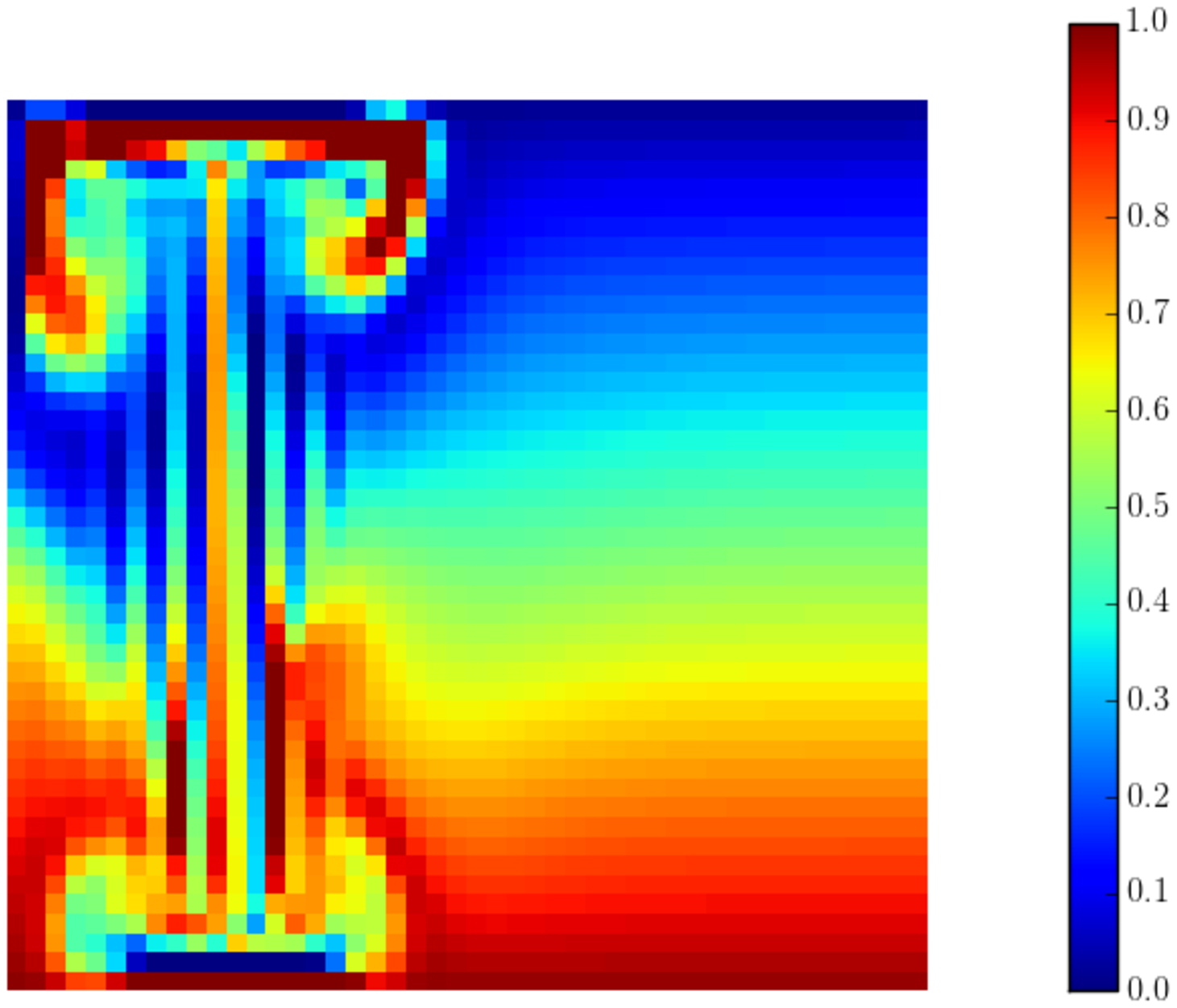}
                  \label{sf:FEC50}
	}
	\\
	\subfigure[\BEC{} scheme]{
	\includegraphics[scale=0.5, trim=70 30 160 30, clip=true]
                  {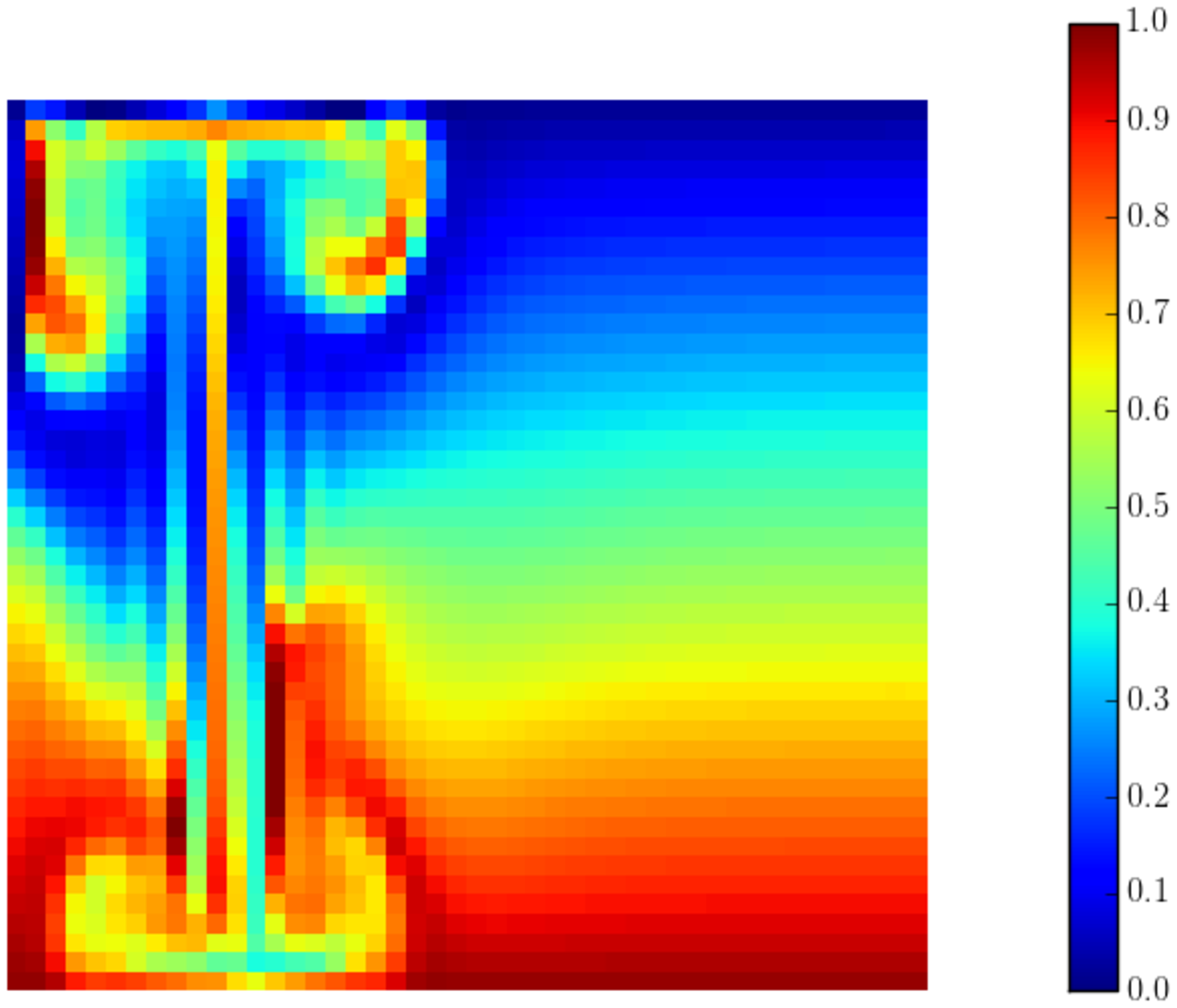}
                  \label{sf:bec50}
	}
	\subfigure[\CEC{} scheme]{
	\includegraphics[scale=0.5, trim=70 30 80 30, clip=true]
                  {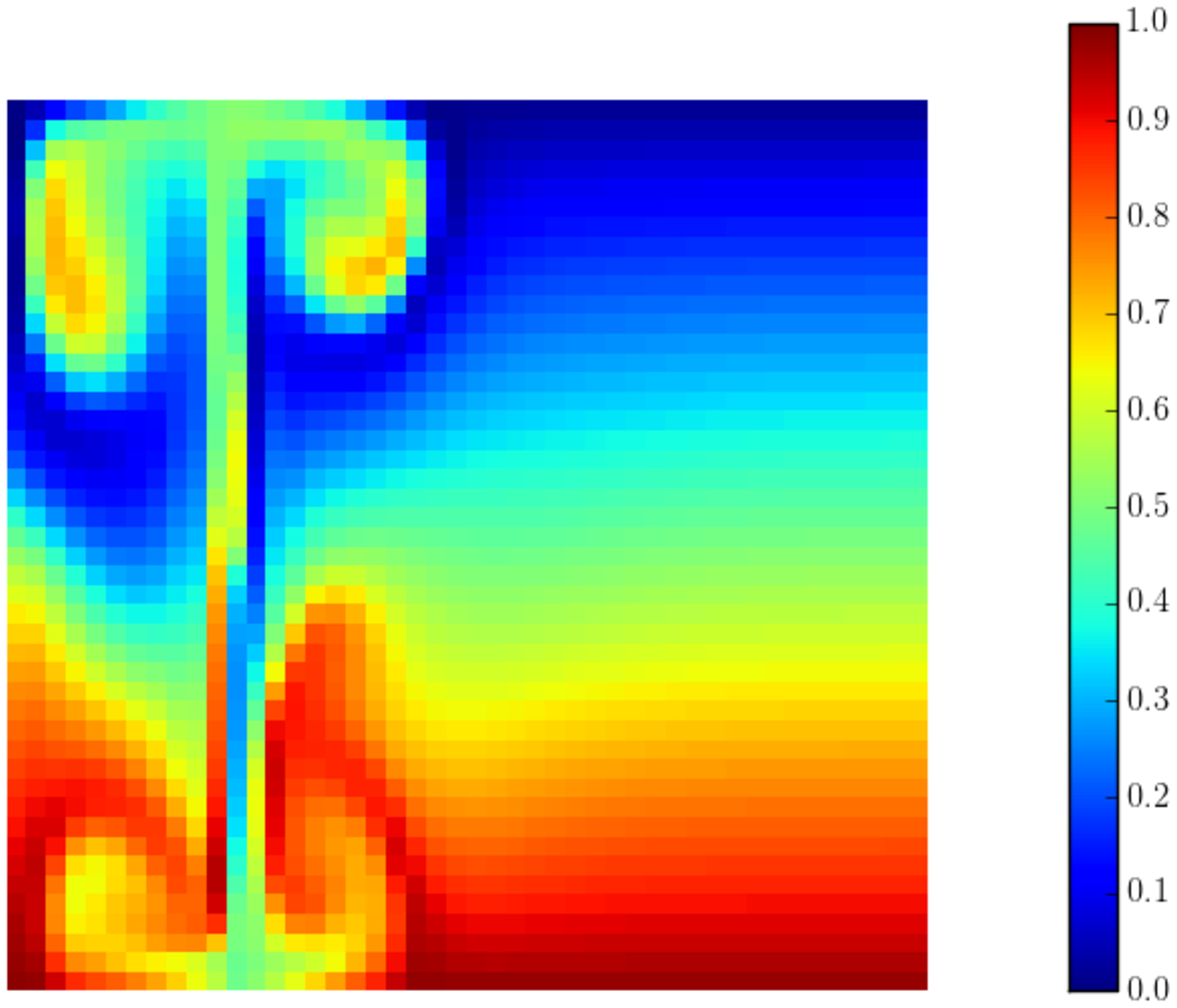}
                  \label{sf:CEC50}
	}
	\caption{Rayleigh-B\'enard evolution of a localized thermal
          perturbation. The numerical resolution $N=50^2$ is intentionally
          modest, in order to highlight numerical errors. }
        \label{f:50}
\end{figure}

In this section, the comparison between the different
advection schemes is extended to a physically more relevant case: 
thermal convection in a layer of fluid heated from below. This
canonical example is also known as the
Rayleigh-B\'enard setup. The schemes will not only be used on passive
scalars that do not influence the velocity, but on the velocity itself and
the temperature, which, in the Rayleigh-B\'enard instability,
modifies the velocity actively. 

The system of equations describing the evolution of the velocity
\U ~and the temperature $T$ of the fluid
is solved on a two-dimensional 
Cartesian domain of aspect ratio $\chi = L_z/L_x = 0.5$, bounded by solid and impermeable
walls. 
The bottom and top plates are maintained at fixed temperatures
$T_0$ and $T_0-\Delta T$, respectively, whereas the vertical walls are
assumed to be insulating (no heat flux through the
vertical boundaries). 
Gravity is assumed to be uniform and vertical~$\gG=-g\ez~.$

To retain the essential physics with a minimum
complexity, the Boussinesq approximation is used to describe the
fluid within the cell and assume that variations of all physical
properties other than density can be ignored. Variations in density
are also neglected ``except in so far as they modify the action of
gravity'' \cite{rayleigh1916}. The density~$\rho$ is assumed to be
constant everywhere in the governing equations except in the buoyancy force
where it is assumed to vary linearly with temperature,
	$
	\rho\left(T\right) = \rhozero \left( 1 - \alpha \left( T -
        \Tzero \right)\right) \,,		
	$
where $\alpha$ is the thermal expansion coefficient of the
fluid. 

The system admits the stationary diffusive solution:
$\U^\star=0$, $T^\star = T_0 - z \, \Delta T / L_z$, and $\grad
P^\star = -g \rho\left(T^\star\right) \ez$. Subtracting the 
stationary solution, choosing $L_z$,
$L_z^2/\kappa$, and $\Delta T$ as units of length, time, and
temperature, respectively, and using the temperature perturbation 
$\theta = T-T^\star$, the system can be written \cite{chandrasekhar1961} as
\begin{align}
	\partial_t \U+ \left(\U\cdot \grad \right)
        \U  
        &= -\grad \Pi + \Ra \Prandtl \theta \, \ez 
	 + \Prandtl \laplacien \U \,,  \label{e:nsnd}\\
	\partial_t \theta 
        + \left(\U\cdot \grad \right) \theta 
        &= w + \laplacien \theta \,, \label{e:Tnd}\\
	\diV \U &= 0 \label{e:mnd} 
	\,,
\end{align}
with $w\equiv\U\cdot \ez$ the vertical velocity.
The non-dimensional control parameters are
the Rayleigh number, defined by $\Ra = \alpha g \Delta T
L_z^3/(\kappa \nu) $ and which measures the convective driving, and
the Prandtl number, defined as the ratio of viscous to thermal
diffusion, $\Prandtl = \nu/\kappa $, with $\nu$ the kinematic viscosity,
$\kappa$ the thermal diffusivity. 

Equations~\eqref{e:nsnd} and \eqref{e:Tnd} are discretized on a
uniform grid using finite volume formula of order two in space and
order one in time, with all the terms being treated explicitly.  To
enforce the solenoidal constraint \eqref{e:mnd}, the pressure-correction 
scheme \cite{chorin1968,guermond2006} is used. This
splitting method is composed of two steps. In the first step, a
preliminary velocity field $\U^{\star}$ is computed by neglecting the
pressure term in Navier-Stokes equation. Since this preliminary
velocity field is generally not divergence-free, it is then corrected
in a second step by a projection on the space of solenoidal vector
fields. Given the temperature and velocity distributions at time-step
$n$, the velocity $\U^{n+1}$ is computed by solving
\begin{eqnarray}
	\U^{(1)} &=& L\left [ \U ^{n},  \U ^{n} \right ]
	\,, \\
	\U^{(2)}&=& \U^{(1)}
		+ \Delta t \left( \Ra \Prandtl \theta^n \ez
		+ \left[\laplacien \U\right]^n \right) 
	\,, \\
	\laplacien \phi^n &=& \diV\U^{(2)} 
	\label{e:poisson}
	\,, \\
	\U^{n+1} &=& \U^{(2)} - \grad \phi^n \,. \label{e:proj}
\end{eqnarray}

In \eqref{e:poisson}, the algorithm
requires to solve at each time-step a Poisson equation for the pressure. The
necessary impermeability conditions for the field $\phi$ are found by multiplying
\eqref{e:proj} by the normal vector $\N$. Together with 
the velocity boundary condition, they lead to 
	$
	\N\cdot \grad \phi^n = 0 
	\,.
	$
The boundary conditions for the velocity field are no-slip,
i.e. $\U = 0$, while the temperature satisfies $\theta(z=0)
=\theta(z=1) = 0$ on the horizontal boundaries, and $\partial_x \theta
=0 $ on the vertical boundaries.
Boundary conditions are imposed on the intermediate
velocity field~$\U^\star$ by introducing ghost points outside of
the domain. In consequence, the tangential component of the actual
velocity field $\U$ will not exactly satisfy the boundary
conditions (the error being controlled by the time-step).

In order to develop the instability (the Rayleigh number being 
sufficiently large and the Prandtl number set to unity), the simulations 
were always started with $\U=0$ and with a
small temperature perturbation. This temperature perturbation consisted 
of a hot spot ($\theta = 0.1$) next to
a cold spot ($\theta = -0.1$). This perturbation, localized close to the lower left
corner, generates a rising and a sinking plume.
The different simulations were compared when the rising plume
has reached the top boundary (after roughly a thousand iterations). 

A very low resolution, $N=50^2$, was deliberately chosen in order to highlight 
the numerical errors associated to the different schemes. 
Snapshots of the total temperature $T = T^\star + \theta$ associated with the thermal plume
are compared on figure~\ref{f:50}.
In fig.~\ref{sf:FEC50} and \ref{sf:bec50}, strong
ripples appear in the wake of the plumes. They are not physically relevant
and are characteristics of dispersive schemes. The
comparison of the plumes in fig.~\ref{sf:CIR50} and
fig.~\ref{sf:CEC50} clearly highlights that the \CEC{} scheme is less
diffusive than the \CIR{} scheme for practical physical applications.
The \CEC{} scheme offers an improved scheme, with significantly reduced
diffusive effects, and free of the strong dispersion characterizing the
\FEC{} and \BEC{} schemes.

%%%%% CCL %%%%%
\section{Conclusion}

Using the simplest semi-Lagrangian \CIR{} scheme introduced by
Courant-Isaacson-Rees, it has been demonstrated that a simple
multi-stage approach can increase the order of the scheme from first
to third order. The resulting scheme is, at leading order, non-dispersive. This
procedure was shown to yield significant improvement on a thermal
convection problem. It can easily be used to increase the order of
existing codes on parallel computers, as the communication stencil is
unaltered by the multi-stage approach.  The communications among
parallel processes are then restricted to the strict miminum (one
layer of cell at each domain boundary).

The \CEC{} algorithm, introduced here, only requires a modest increase in the computational cost
and can easily be implement in existing codes.
Moreover, its implementation is not limited to 
regular Cartesian finite differences schemes. It can be 
generalized to other geometries and scheme types by 
following two simple steps:
(i)~deriving the modified advection equation for the \FEC{} and \BEC{} schemes and 
(ii)~combining both schemes to cancel out their leading order error.

%%%%% Biblio %%%%%
\bibliographystyle{wileyj}
\bibliography{hola}

%%%%% APPENDIX %%%%%
\appendix
\section{Developed expressions of the corrective schemes}
	The expressions relevant to \eqref{eq:cstUHALO} and \eqref{eq:cstUBEC} can be developed as
	\begin{align}
		\label{eq:exp:FEC}
		2\, \FEC[\Phi]_i 
		=& - U_i (1-U_i ) \Phi^n_{ i + s_i } 
		+ (2- U_iU_i)\Phi^n_i 
			\\&
		- U_i U_{i+s_i} \Phi^n [ i + s_i - \ign{i+s_i}] 
		+ U_i ( 1 + U_{i-s_i} ) \Phi^n_{i - s_i} \, ,
		\nonumber
	\end{align}
	
	\begin{align}
		\label{eq:exp:bec}
		2\, \BEC [\Phi]_i =& 
		\left(f\Phi^n\right) [ i+ \ign{i}]
		+
		\left(f \Phi^n \right )[i] 
		+
		\left( f \Phi^n \right)
		[i - \ign{i} + \ign{i - \ign{i}}] 
		+
		\\ \nonumber	&
		\left( f 	\Phi^n \right)
		[i + \ign{i} - \ign{i+ \ign{i}}]
		+ 
		\\ \nonumber	&
		\Big[
			\left( f \Phi^n\right)[i - \ign{i}]		
			+
			\left( f \Phi^n \right)
			[i - \ign{i} + \ign{i + \ign{i}} - 
	 			\ign{i - \ign{i} + \ign{i - \ign{i}}}] 
		\Big]
		+
		\\ &\nonumber		
		\left(f \Phi^n\right)
		[i - \ign{i} - \ign{i - \ign{i}}] 
		\, , 
	\end{align}	where
	\begin{align}
				&
		f[ i+ \ign{i} ] = - (1 - U_i) U_i (1 - U_{i+ \ign{i}}) 
			\; , \\&
		f[ i ] = ( 1 - U_i) \big[ 3 - (1 - U_{i})^2 \big]
			\; , \\&
		f[ i - \ign{i} + \ign{i - \ign{i}}]	= 
		- U_{i} U_{i- \ign{i} } ( 1 - U_{i- \ign{i} + \ign{i - \ign{i} }})
			\; , \\&
		f[ i + \ign{i} - \ign{i+ \ign{i}} ] 
		= - ( 1 - U_{i}) U_i U_{i+ \ign{i}}
			\; , \\&
		f[i - \ign{i}] = 
			U_i \big[ 3 \,- \left (1 - U_{i- \ign{i} } \right)^2 \big] 
			-\,
			( 1 - U_i) \big( (1- U_i) U_i \big) 
			\; , \\&
		f [i - \ign{i} + \ign{i + \ign{i}} - 
	 	\ign{i - \ign{i} + \ign{i - \ign{i}}}] 
		=- U_{i} U_{i - \ign{i}} 
	 	U_{i - \ign{i} + \ign{i - \ign{i}}} 
	 		\; , \\&
		f[i - \ign{i} - \ign{i - \ign{i}}] =
		 - U_{i} (1- U_{i - \ign{i} }) U_{i - \ign{i} }
		\; . 
	\end{align}
	
\section{Analysis of the modified advection equation}
The modified equation steming from the discretization of the advection equation has in one dimension
the general form
	\begin{align}
		\partial_t \Phi + u\,\partial_x \Phi = 
		\sum_{\alpha} C_{\alpha} \partial^{\alpha}_{x} \Phi \, , 
		\label{eq:genmodeq}
	\end{align}
	where the $C_{\alpha}$ prefactors come from the truncation error 
	in the case of numeric schemes.
	If the CFL stability condition is met, i.e. $\Delta t \propto u^{-1} \Delta x$, 
	with $\Delta x \propto N^{-1}$, we have
	\begin{align}
		C_{\alpha} \propto N^{-\alpha+1} \, . 
		\label{eq:coefres}
	\end{align}
	Going into Fourier space for spacial dimensions 
	and Fourier-Laplace space for time,
	\begin{align}
		\Phi(x,t) = \int \ddif k \;
		{\rm e}^{ \Omega (k) t - i k x }\hat{{\Phi}} ( k, & \Omega (k) ) 
			\quad \text{where} \quad 
		\Omega (k) = -\sigma (k) + i \omega (k) \,.
		\label{eq:fourTA}
	\end{align}
	Thus, the dispersion relation is
	\begin{align}
		\Omega (k)= (ik) u + \sum_{\alpha} (-ik)^{\alpha}C_{\alpha} \,. 
	\end{align}
\label{adix:Meq:stab}
	Using the decomposition introduced in \eqref{eq:fourTA},
 	the decay rate and the phase drift can be expressed as
		\begin{align}
			\sigma (k) &= \sum_{p} 
			\left( k^2 \right) ^{2p+2} \left ( C_{4p+2} 
			\!-\! 
			\left( k^2 \right)^{2p} C_{4p}  \right ) \, ,\\
			\omega (k) &= k\Big( u - \!
			\sum_{p} \left( (k^2)^{2p} C_{4p+1} 
			\!-\!
			(k^2)^{2p+1} C_{4p+3} \right) \Big)
				\, .
		\end{align}
	The equation has strictly stable solutions if and only if $\sigma(k) > 0$. 
	Because of their dependence on the resolution, the sequence of $C_{2p}$
	is often equivalent to its first term different from zero. The stability 
	reduces to the criterion $ C_{\alpha} > 0$ if $\alpha = 4p+2$ and
	$ C_{\alpha} < 0 $ if $\alpha = 4p$.
	Using the equation on $\omega$, the phase drift can be extracted
	\begin{align}
		\phi(k) = \omega (k) - ku = -k \sum_{p} \!\! \left(  (k^2)^{2p} C_{4p+1}
			-
			(k^2)^{2p+1} C_{4p+3} \right) \, .
		\label{eq:phacoef}
	\end{align}	
\label{adix:Meq:rev}
It is important to note that the procedure introduced in the \FEC{} scheme cannot be repeted recursively.
In order to highlight this point let us note that for pure advection, reversing time
is equivalent to reversing the velocity
	\begin{align}
		\partial_{-t} \Phi + u \partial_x \Phi = 0 
		\quad \Leftrightarrow \quad 
		\partial_{t} \Phi + (-u) \partial_x \Phi = 0 
		\quad \Leftrightarrow \quad 
		\partial_{t} \Phi + u \partial_{- x} \Phi = 0 \,.
	\end{align}
	Going into Fourier space for the spacial dimension
	\begin{align}
		\Phi(x,t) = \int \ddif k \,
			{\rm e}^{ - i k x }\tilde{\Phi} ( k,  t) \, ,
		\label{eq:four}	
	\end{align}
	the modified advection equation can be written as
	\begin{align}
		\partial_t \left( \ln \tilde{\Phi} \right)( k,  t) 
		= u(ik) + \sum_{\alpha} C_{\alpha}(-ik)^{\alpha} \, .
	\end{align}
Reversing the sign of the coordinate, $x \!\rightarrow\! - x$, is equivalent to reverse the
wave vector, $k \!\rightarrow\! - k $ (c.c. for a real field). 
In order to ensure time reversibility, the following relation should be satisfied
	\begin{align}
		\partial_t \left( \ln \tilde{\Phi} \right) ( k,  t) 
		&= \partial_t \left( \ln \tilde{\Phi} \right) ( - k, -t)
		=- \partial_t \left( \ln \tilde{\Phi} \right) ( - k,  t) \,.
	\end{align}		
	This last relation shows that only terms of odd derivative are reversible.
	The error on $\bar{\Phi}$ highlights this observation.
	It can be evaluated using
	\begin{align}
		\left( \ln \tilde{\bar{\Phi}} \right) ( k,  t ) 
		&=
		\left( \ln \tilde{\Phi} \right) ( k,  t ) 
		+
		2 \Delta t \sum_{p} C_{2p}(ik)^{2p}.
	\end{align}
	Only terms of even order derivative modify the field and can be detected with this procedure. 
	This property should also be true for the $C_{\alpha}$
	coefficients when the velocity is reversed.
	In the case of the \CIR{} scheme, the coefficients depend 
	on the sign of the velocity.
	In the case of the non-ideal advection equation \eqref{eq:genmodeq}, 
	reverting time leads to
	\begin{align}
		\partial_{t} \Phi + (- u)\partial_x \Phi 
		&=
		\sum_{p} \! \Big ( 	
		C_{2p+1}(- u) \partial^{2p+1}_{x} \Phi 
			-
		C_{2p} (- u)\partial^{2p}_{x} \Phi 
		\Big )
		\, . 
	\end{align}	
	Once more, only terms of odd order derivative are reversible. 

\begin{figure}[H]
	\subfigure[Decay rate]{
	\includegraphics[scale=0.4,trim= 20 10 50 40, clip=true]{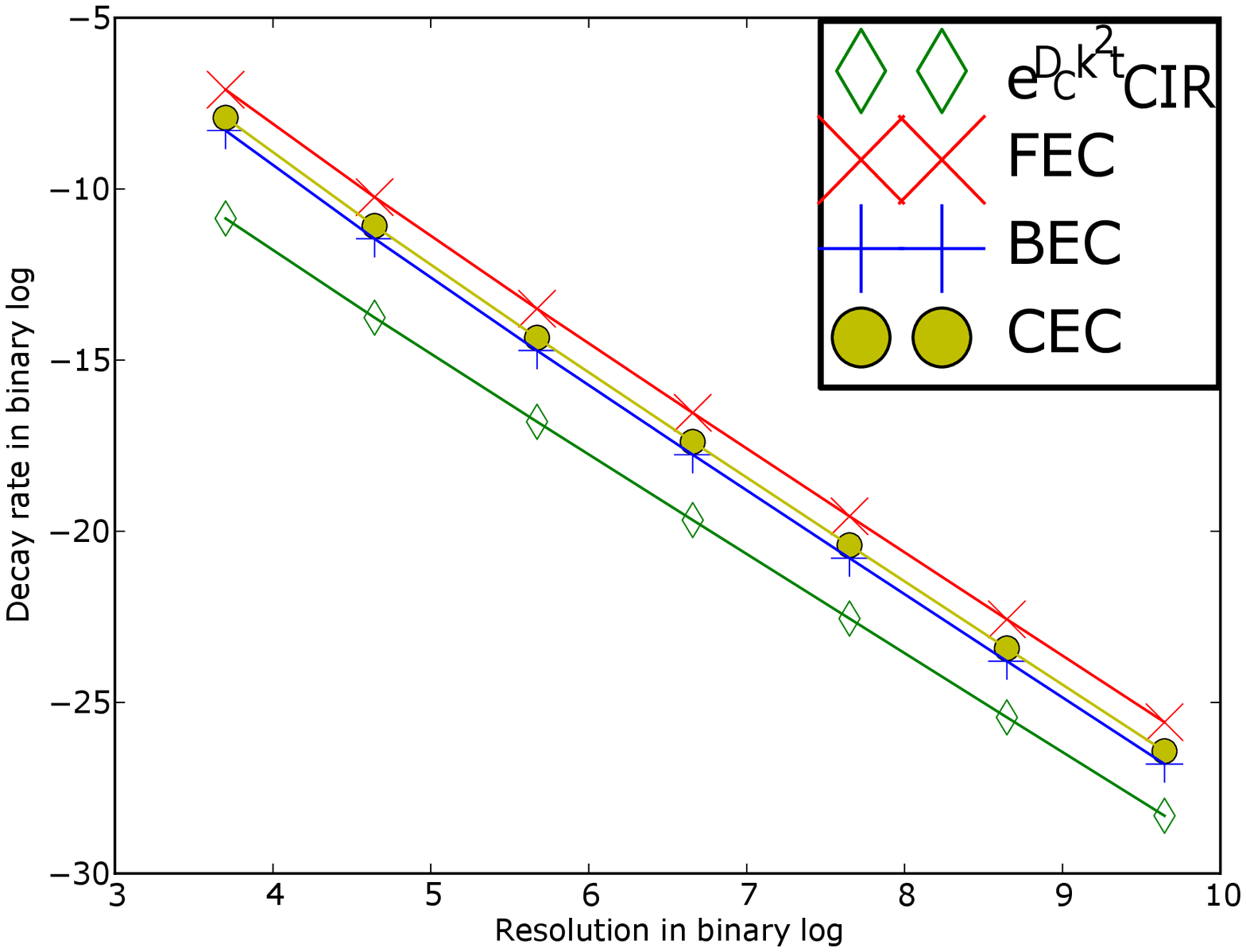}
	\label{fig:grw}	 }
	\subfigure[Rescaled decay rate]{
	\includegraphics[scale=0.4,trim= 20 10 50 40, clip=true]{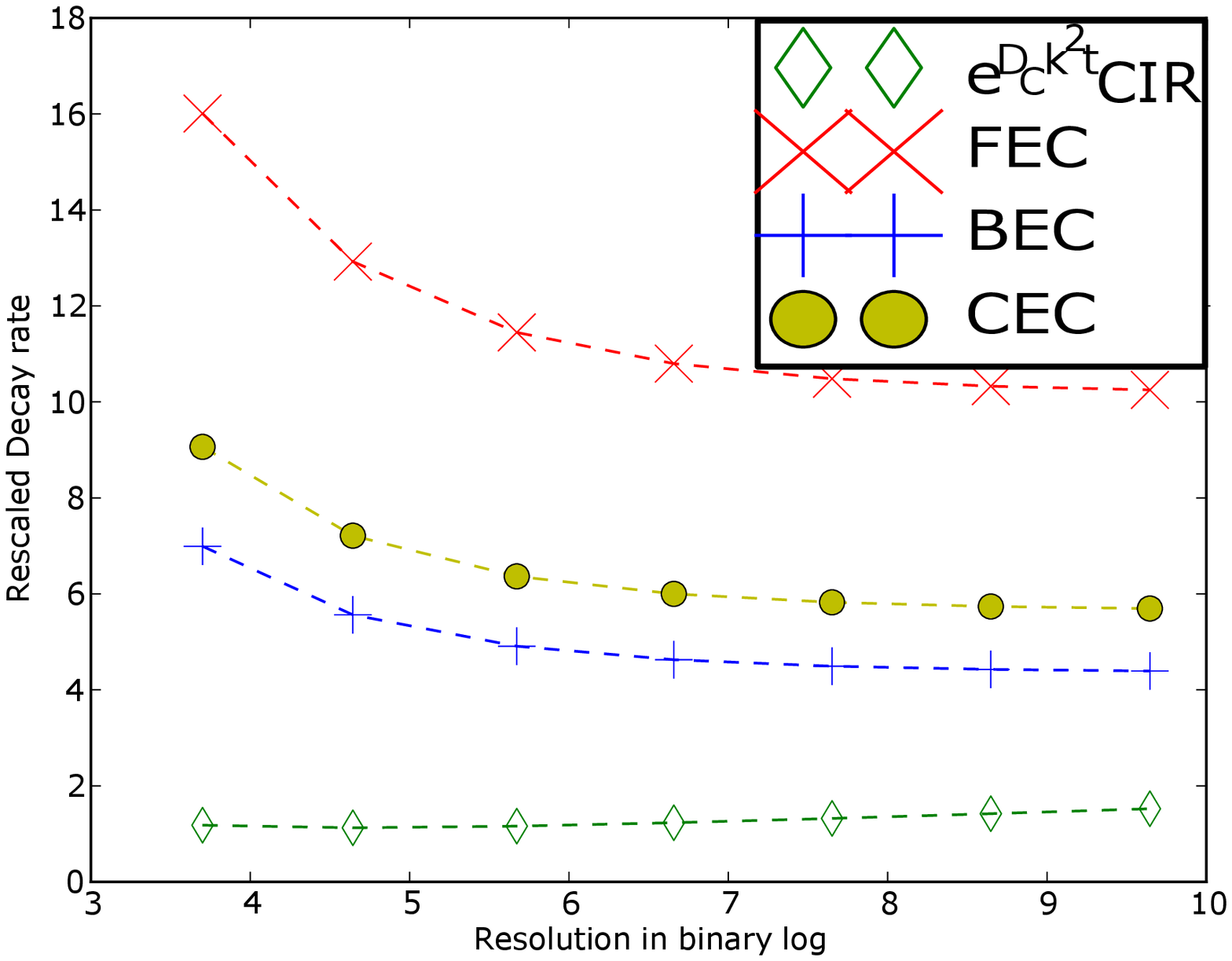}
	\label{fig:grwR} }
	\caption{Evolution of the decay rate with the resolution in one dimension.}
	\label{fig:1DresplotDR}
\end{figure}

\begin{figure}[H]
	\subfigure[Phase drift]{
	\includegraphics[scale=0.4,trim= 20 10 50 40, clip=true]{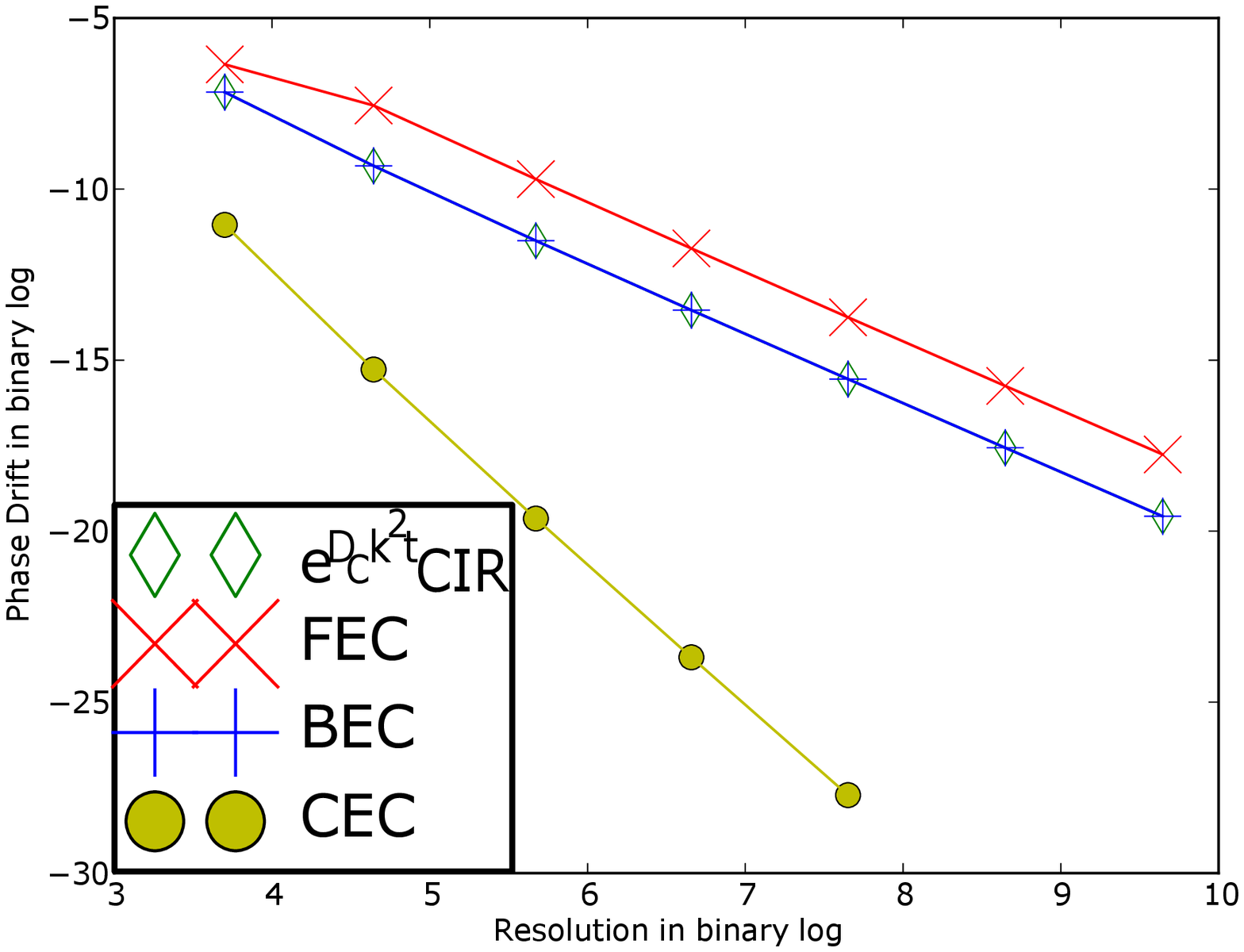}
	\label{fig:pha}	 }
	\subfigure[Rescaled phase drift]{
	\includegraphics[scale=0.4,trim= 20 10 50 40, clip=true]{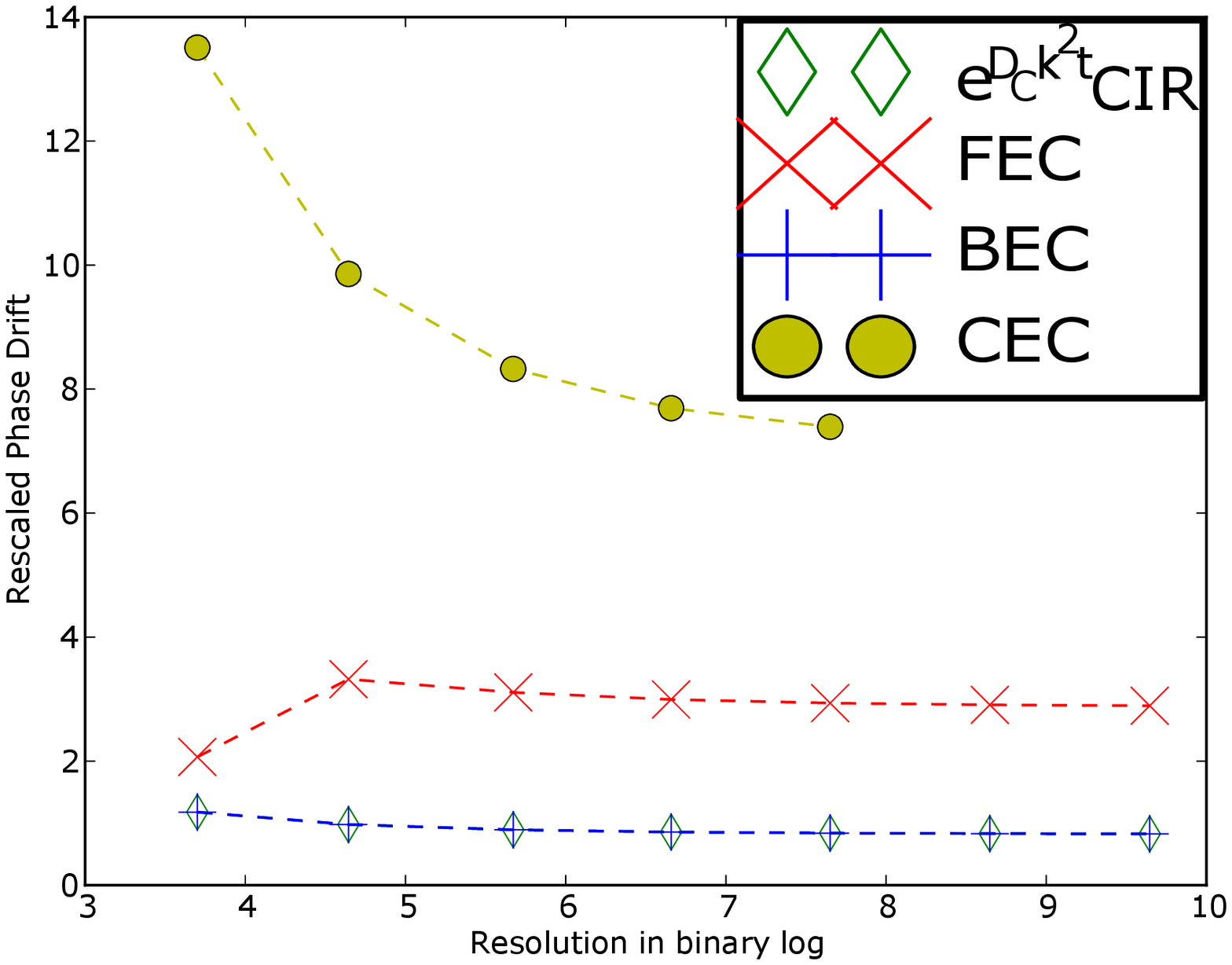}
	\label{fig:phaR} }
	\caption{Evolution of the phase drift with the resolution in one dimension.}
	\label{fig:1DresplotPD}
\end{figure}

The decay rate (fig.~\ref{fig:1DresplotDR}) and the phase drift (fig.~\ref{fig:1DresplotPD}) 
	were 
	measured for different resolutions. 
	The results are plotted as a 
	function of the resolution on a binary log scale ($\lb$). fig.~\ref{fig:grw} and \ref{fig:pha} represent the decay rate and the phase drift, respectively. 
	As shown in \eqref{eq:coefres}, the prefactors of the derivative terms of the error
	are proportional to an integer power of the resolution, $C_{\alpha} \propto N^{-\alpha+1}$. 
	The values of $\alpha$ are in good agreement with the error term of the 
	modified equation. Using the theoretical value of $\alpha^{(1)}$ and $\alpha^{(2)}$, 
	the values are rescaled to
	$
			\phi_{res} = \phi \!\times\! N^{\alpha^{(1)}-1} 
	$ and
	$
			\sigma_{res} = \sigma \!\times\! N^{\alpha^{(2)}-1} \, .
	$
	fig.~\ref{fig:grwR} and \ref{fig:phaR} show that the rescaled values 
	are nearly constant as predicted by the theory. 

%%%%%%% END
\end{document}